\documentclass[10pt,aps,prb,twocolumn,superscriptaddress,longbibliography,amsmath,amssymb]{revtex4-2}

\usepackage{graphicx}
\usepackage{bm}
\usepackage{dcolumn}
\usepackage{listings}
\usepackage{xcolor}
\usepackage[utf8]{inputenc}
\usepackage{hyperref}
\DeclareUnicodeCharacter{2194}{\leftrightarrow}
\begin{document}

\title{The electric field gradient tensor as a symmetry-adapted order parameter in Landau theory}

\author{L. Scalise}
\affiliation{Instituto de Pesquisas Energ\'eticas e Nucleares, IPEN/CNEN, S\~ao Paulo, Brazil}

\author{A. W. Carbonari}
\affiliation{Instituto de Pesquisas Energ\'eticas e Nucleares, IPEN/CNEN, S\~ao Paulo, Brazil}

\date{\today}

\begin{abstract}
Quadrupolar hyperfine spectroscopies, Nuclear Quadrupole Ressonance (NQR), Nuclear Magnetic Ressonance (NMR), Time-Differential Perturbed Angular Correlations (TDPAC), and M\"ossbauer, have long used the electric field gradient (EFG) at a nuclear site as an empirical proxy for order parameters at structural and electronic phase transitions, but the EFG itself has never been given a systematic role in Landau theory. Here we supply that construction. The EFG is an exactly traceless, symmetric, pure rank-2 tensor defined at a crystallographic site; decomposing it under the site-symmetry group and inducing over the Wyckoff orbit yields its content in irreducible representations of the parent space group. Whenever a zone-center transition's representation appears in this content, symmetry forces the corresponding EFG combination to vanish above the transition and grow linearly in the order parameter below it, inheriting its critical exponent, sign, and domain structure: it is a symmetry-adapted realization of the
order parameter. Complementary channels grow quadratically, deriving the classic empirical relations. The resulting primary/secondary/forbidden classification is falsifiable and is validated in two independent halves: temperature dependence by five decades of quadrupolar measurements, from critical-exponent extractions to first-order discontinuities and one
exclusionary null result, and structural form by first-principles calculations obeying the framework's parity and zero theorems. All-electron calculations on $\alpha$-quartz confirm the orbit-selection component of this structural prediction: among electric field gradient combinations built from one crystallographic orbit, only the combination matching the soft-mode irrep switches on linearly with the distortion amplitude, while its symmetry-orthogonal partner stays suppressed by two orders of magnitude. A definitive test of the complementary single-site parity and zero rules is identified and left to future work. A proposed case study of the $^{75}$As site through the nematic transition of BaFe$_2$As$_2$ sets out five falsifiable predictions, including a previously unstated null.
\end{abstract}

\maketitle

\section{Introduction}\label{sec:intro}

Continuous phase transitions are organized by a single powerful idea: below the transition, the system is described by an order parameter---a quantity that vanishes identically in the high-symmetry phase, grows continuously in the low-symmetry phase, and transforms according to an irreducible representation of the parent symmetry group \cite{2013Landau,1987Toledano}.
The choice of \emph{which} physical quantity to promote to order-parameter status is, however, less rigid than textbook presentations suggest: the magnetization, the spontaneous polarization, the soft-mode coordinate, or the spontaneous strain may all serve for one and the same transition, because Landau theory fixes only the representation, not its physical
carrier. Historically, each newly recognized carrier has opened an experimental window---the spontaneous strain gave ferroelastics their quantitative Landau theory \cite{1990Salje,1998Carpenter}, and the Landau--de Gennes tensor gave nematic liquid crystals theirs
\cite{1993DeGennes}. This paper argues that a further carrier has been overlooked, one that is measured routinely, site-selectively, and with exquisite precision in solids: the electric field gradient (EFG) tensor at a nuclear site.

The EFG---the traceless symmetric part of the second spatial derivatives of the electrostatic potential at a nucleus---is the coupling partner of the nuclear quadrupole moment, and is measured by nuclear quadrupole resonance (NQR), quadrupole-perturbed nuclear magnetic resonance (NMR), perturbed angular correlation (PAC), and M\"ossbauer spectroscopy \cite{1996Schatz}. Its virtues as a probe of phase transitions have been appreciated for half a century: quadrupolar frequencies shift and lines split at structural transitions, and the classic NMR--NQR literature established empirical relations of the form $\Delta\nu_Q \propto \langle\varphi\rangle$ or $\langle\varphi^2\rangle$ between quadrupolar observables and the order parameter $\varphi$ \cite{1979Borsa,1984Rigamonti}. On the theoretical side, the EFG is now computed routinely and accurately from first principles by all-electron density functional methods \cite{1988Blaha, elk, wien2k}. Yet throughout this literature---experimental and computational alike---the EFG appears in a supporting role: a spectator quantity coupled to an order parameter defined elsewhere (a displacement, a polarization, a nematic amplitude), and almost always reduced to two scalars, the principal component $V_{zz}$ and the asymmetry parameter, with the remaining degrees of freedom of the tensor discarded.

We contend that this casting understates what the EFG is. Three observations, developed in this paper, elevate it from spectator to principal. First, the EFG is \emph{exactly} a rank-2 traceless symmetric tensor---a pure $\ell=2$ object---defined at a crystallographic site: it
therefore decomposes, by subduction onto the site-symmetry group and induction over the Wyckoff orbit, into irreducible representations of the parent space group, which is precisely the structure Landau theory demands of an order parameter. Second, whenever a transition's irrep appears in this decomposition, symmetry forces the corresponding EFG combination to vanish identically above $T_c$ and to grow linearly in the order parameter below it, with the same critical exponent---making that combination a \emph{bona fide realization of the order parameter}, in the same equivalence-class sense in which the spontaneous strain realizes a
ferroelastic order parameter. Third, the complementary case---EFG combinations belonging to other representations---grow quadratically, reproducing the classic $\Delta\nu_Q\propto\varphi^2$ phenomenology as a theorem rather than an ansatz. The resulting linear/quadratic dichotomy is falsifiable: for a given space group, Wyckoff position, and transition channel, group theory alone dictates which quadrupolar observables inherit
the order-parameter exponent $\beta$ and which show $2\beta$, which must remain identically zero, and which merely shift smoothly.

To our knowledge, no systematic treatment of the EFG tensor as a symmetry-adapted order parameter exists. The empirical order-parameter behavior of quadrupolar observables is well documented---from the critical-exponent extraction $\beta = 0.250\pm0.005$ by $^{35}$Cl NQR in a
layered perovskite \cite{1976Kind}, through the two-exponent critical analysis of incommensurate Rb$_2$ZnCl$_4$ by $^{87}$Rb satellite NMR \cite{1982Schneider}, to the recent identification of Landau-type order-parameter behavior of NMR splittings at nematic transitions in iron-based superconductors \cite{2015Bohmer,2015Wiecki}---but in every case the symmetry analysis was performed on the presumed microscopic order parameter, with the EFG deduced from it case by case. Site-symmetry constraints on EFG tensors are textbook material and are applied compound
by compound \cite{2026Derevianko}, but no general framework connects them to the Landau free energy; and while Landau theories of rank-2 tensor order parameters are mature for liquid crystals and electronic nematics \cite{1993DeGennes,2014Fernandes}, the tensor in those theories is never the nuclear-site EFG itself. While numerous studies have demonstrated empirical correlations between electric field gradients and structural order parameters, these correlations have generally been interpreted phenomenologically, with the EFG regarded as a secondary observable responding to an independently defined order parameter. What has remained absent is a symmetry-based criterion capable of determining a priori whether a given EFG component constitutes an independent realization of the broken symmetry itself. The present work addresses precisely this gap by combining site-symmetry representation theory, induction over Wyckoff orbits, and Landau free-energy analysis into a unified framework that assigns each EFG channel to one of three symmetry classes: primary, secondary, or forbidden. The gap, precisely stated, is the absence of the construction that this paper supplies: the irreducible decomposition of the EFG at arbitrary Wyckoff positions, the Landau free energy written directly in EFG variables, and the resulting predictive classification of quadrupolar observables at symmetry-breaking transitions.

This paper provides that construction and validates it. Section~\ref{sec:theory-efg} establishes the mathematical structure of the EFG---symmetry, tracelessness, and the resulting five independent components---with care taken at one point where the standard argument is an
idealization: tracelessness at a real nuclear site follows not from Laplace's equation, which fails in the presence of finite contact density, but from the exact decoupling of the trace from the quadrupole interaction. Sections~\ref{sec:theory-efg} and~\ref{sec:sym-adapted} state the Landau criterion for zone-center transitions, the case treated throughout, and develop the required representation theory, including the step most often elided: the EFG transforms under the \emph{site} group, while the Landau criterion refers to the \emph{parent} group, and the bridge is the representation induced over the Wyckoff orbit. Section~\ref{sec:sym-adapted} derives the transformation properties explicitly, proves the vanishing of the EFG at cubic sites as the sharp special case of a general counting rule, and establishes the linear/quadratic coupling dichotomy. Section~\ref{sec:efg-as-op} proves the central claim---the conditions under which an EFG combination is a realization of the order parameter---and delimits it honestly: the EFG realizes the order parameter; it does not drive the transition. Section~\ref{sec:free-energy} constructs the Landau free energy in EFG variables, recovering the Landau--de Gennes theory as the isotropic limit and deriving its channel-resolved crystalline generalization, including a strain--EFG cross-coupling that predicts a Curie--Weiss divergence of the strain-induced EFG response above $T_c$---a proposal we term strain-NQR. Sections~\ref{sec:expt} and~\ref{sec:dft} confront the framework with the experimental and first-principles records respectively, organized by the dichotomy: the two halves of the Landau prediction---the symmetry-dictated structure of the EFG--order-parameter map, and the thermodynamic behavior of the order parameter itself---are validated by DFT and by experiment separately, and compose without adjustable structure. Section~\ref{sec:case} assembles the complete argument for a single material, BaFe$_2$As$_2$, where the $^{75}$As site symmetry, the $B_{2g}$ transition channel, the orbit induction, the free energy, the spectroscopic dictionary, and new channel-resolved DFT combine into five falsifiable signatures, including one null prediction that has, to our knowledge, not been stated before.

Beyond its conceptual point, the framework has a practical payoff that motivated this work: it makes first-principles hyperfine calculations and Landau phenomenology mutually calibrating. The coupling coefficients of the EFG free energy are exactly the derivatives that frozen-distortion DFT calculations compute; conversely, the symmetry-dictated parity and zero structure of those calculations furnishes internal consistency checks that the DFT-EFG literature has not exploited. For the growing program of using hyperfine parameters as quantitative probes of electronic structure, topology, and phase transitions, the present work supplies what has been
missing: the theorem that says precisely \emph{when} a quadrupolar observable is the order parameter, and when it is only its shadow.

Density-functional calculations provide an ideal platform for testing the symmetry classification proposed here because the vanishing of forbidden channels is independent of the exchange-correlation functional. In contrast, the numerical values of the coupling coefficients are expected to retain the usual functional dependence characteristic of EFG calculations. Accordingly, the present framework distinguishes between symmetry-protected predictions, which are exact, and quantitative amplitudes, which remain subject to the approximations of electronic-structure theory.

Section~\ref{sec:dft-quartz} then verifies the construction directly by first-principles calculation on $\alpha$-quartz, where the primary, secondary, and forbidden channels appear together and their parity, leading power, and > exact zeros are confirmed.

\section{Theory}\label{sec:theory}

\subsection{Mathematical structure}\label{sec:theory-efg}

We consider a nucleus at the origin, embedded in the electrostatic potential $V(\mathbf r)$ generated by all charges external to it. \emph{Assumption A1 (static sources):} the charge distribution is static on the timescale of the quadrupole interaction; $V$ is the thermally averaged Born--Oppenheimer expectation value, and fluctuations enter relaxation, not the static splittings treated here. \emph{Assumption A2 (regularity):} $V\in C^2$ near the origin, guaranteed by any bounded, continuous electronic density.

The second-derivative tensor is $V_{ij}\equiv\partial_i\partial_j V|_0$. Taylor-expanding $V$ against the nuclear density $\rho_N$, \begin{equation}
E = ZeV(0) + \sum_i V_i\!\!\int\!\rho_N x_i
  + \tfrac{1}{2}\sum_{ij}V_{ij}\!\!\int\!\rho_N x_i x_j + \dots,
\end{equation}
the monopole term is orientation-blind and the dipole term vanishes by nuclear parity. Decomposing $x_ix_j$ into traceless and trace parts, the trace couples only to the nuclear mean-square radius (the isomer-shift channel), while the orientation-dependent energy is
\begin{equation}
E_Q = \tfrac{1}{6}\sum_{ij}Q_{ij}V_{ij},\qquad
Q_{ij}\equiv\!\int\!\rho_N(3x_ix_j - r^2\delta_{ij}),
\label{eq:EQ}
\end{equation}
with $Q_{ij}$ traceless and symmetric by construction; hence only the traceless symmetric part of $V_{ij}$ is observable in $E_Q$.

\emph{Symmetry.} By A2 and Schwarz's theorem, $V_{ij}=V_{ji}$: a consequence of regularity, not an independent postulate. Components: $9\to6$.

\emph{Tracelessness---the honest statement.} The familiar argument invokes Laplace's equation in a charge-free region; at a real nuclear site this premise fails, since Poisson's equation gives
$\mathrm{Tr}\,V = -\rho_e(0)/\varepsilon_0 \neq 0$ from the finite $s$-electron contact density. The rigorous route is the one already prepared: (i) the trace decouples exactly from every quadrupolar observable; (ii) one therefore \emph{defines} the EFG as the traceless projection,
\begin{equation}
V^{\mathrm{EFG}}_{ij}\equiv \partial_i\partial_jV|_0
 - \tfrac{1}{3}\delta_{ij}\nabla^2V|_0,
\end{equation}
the quantity computed by all-electron codes and measured by quadrupolar spectroscopies. With this definition $\mathrm{Tr}\,V^{\mathrm{EFG}}=0$ exactly, with no charge-free idealization. Hereafter $V_{ij}$ denotes $V^{\mathrm{EFG}}_{ij}$. Counting: six symmetric components minus one trace constraint gives \textbf{five independent components}.

\emph{Tensor character.} Under rotations, $V'_{ij}=R_{ik}R_{jl}V_{kl}$ [Supplemental Material (SM), Sec.~S4 \cite{SM}]: a Cartesian rank-2 tensor. Of the $\mathrm{SO}(3)$ content $9 = 1\oplus3\oplus5$, the $\ell=1$ part vanished with the antisymmetric part and the $\ell=0$ part was projected out by definition: \emph{the EFG is a pure $\ell=2$ object}, parity-even and time-even. This purity is the hinge of all that follows.

\emph{Principal axes.} The spectral theorem diagonalizes $V$ to $(V_{XX},V_{YY},V_{ZZ})$ with $V_{XX}+V_{YY}+V_{ZZ}=0$; with the ordering convention $|V_{ZZ}|\ge|V_{YY}|\ge|V_{XX}|$ and the asymmetry parameter $\eta\equiv(V_{XX}-V_{YY})/V_{ZZ}$,
\begin{equation}
V_{XX}=-\tfrac{V_{ZZ}}{2}(1-\eta),\qquad
V_{YY}=-\tfrac{V_{ZZ}}{2}(1+\eta),
\label{eq:pas}
\end{equation}
and the ordering forces $0\le\eta\le1$ (SM, Sec.~S4). The bookkeeping that prefigures the paper: $5 = 2+3$---two shape parameters $(V_{ZZ},\eta)$ plus three PAS Euler angles. Standard practice retains the two and discards the three; at a symmetry-breaking transition the discarded  orientational components can be the critical ones, so the correct order-parameter
candidate is the full five-component object.

\begin{figure}[t]
\includegraphics[width=\columnwidth]{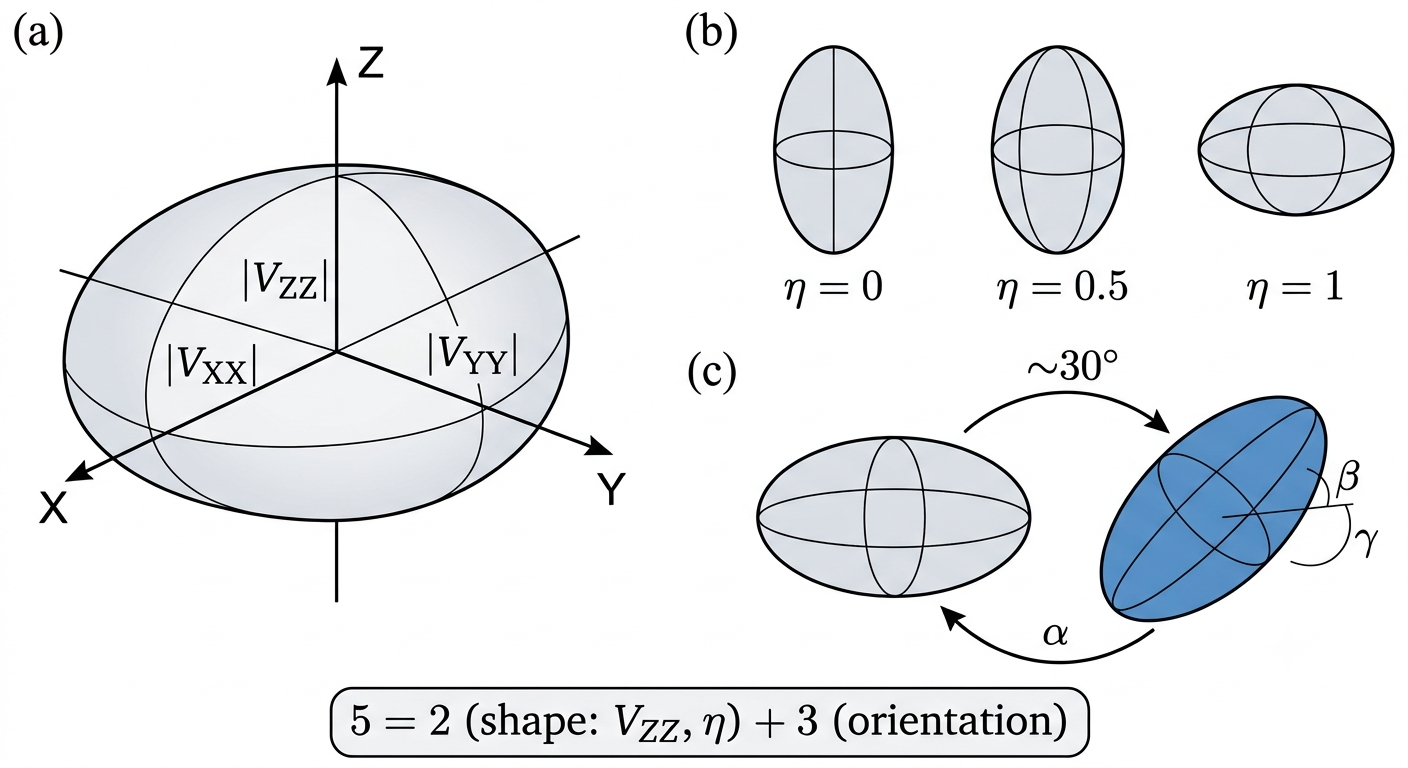}
\caption{Geometric content of the EFG tensor. (a) Representation ellipsoid of the eigenvalue magnitudes in the principal axis system (PAS); the ordering convention $|V_{XX}|\le|V_{YY}|\le|V_{ZZ}|$ fixes the axis labels. (b) The two shape parameters: axial symmetry at $\eta=0$ and maximal biaxiality at $\eta=1$ span the allowed range $0\le\eta\le1$. (c) The three orientational parameters (Euler angles of the PAS), discarded in the conventional $(V_{ZZ},\eta)$ reduction; at a symmetry-breaking transition these can be the critical components, motivating the full five-component treatment.}
\label{fig:ellipsoid}
\end{figure}

Let $G_0$ be the parent symmetry group and $G\subset G_0$ the daughter's. The free energy is a scalar, $F[g\circ\varphi]=F[\varphi]$ for all $g\in G_0$. (i)~Group operations act linearly on the order-parameter components; closure makes the matrices $D(g)$ a representation of $G_0$.
(ii)~If $D$ is reducible, Schur's lemma forbids invariant bilinears between inequivalent irreps and the quadratic term block-diagonalizes with independent coefficients $a_\mu(T)$. (iii)~A continuous transition occurs where one $a_\mu$ changes sign; simultaneous vanishing is a multicritical coincidence, excluded generically (\emph{Assumption A3}).

\emph{Criterion.} A quantity is a Landau order parameter iff its components transform as a single (physically) irreducible representation of $G_0$ and vanish identically in the parent phase. Complex-conjugate irrep pairs combine into real form; the Landau and Lifshitz conditions constrain which irreps admit continuous transitions but not the criterion itself.

\emph{Assumption A4 (scope).} We restrict to zone-center ($\mathbf k=0$, translationengleiche) transitions, for which space-group irreps reduce to irreps of the crystallographic point group $\bar G_0$. Zone-boundary and incommensurate transitions require the induced-representation machinery of Sec.~\ref{sec:sym-adapted} applied to enlarged orbits and are deferred.

\subsection{Representation theoretical framework}\label{sec:sym-adapted}

A symmetry-adapted basis of a space $W$ of physical quantities block-diagonalizes the $G_0$ representation into irreps; the construction is algorithmic via projectors $\hat P^{(\mu)}_{mm}=\frac{d_\mu}{|G_0|}\sum_g [D^{(\mu)}_{mm}(g)]^*\hat O(g)$ and character multiplicities $n_\mu = |G_0|^{-1}\sum_g\chi^{(\mu)}(g)^*\chi_W(g)$ \cite{1972Bradley} (SM, Sec.~S3). For displacements this yields the tabulated symmetry modes; for strain, spontaneous-strain theory; for the EFG carrier space, the present construction.

\emph{The site/parent bridge.} The EFG lives at a site $\mathbf q$ with site group $G_{\mathbf q}\subset G_0$. (a)~For $g\in G_{\mathbf q}$ the single-site tensor transforms into itself; Neumann's principle then annihilates, in equilibrium, every component outside the identity irrep of $G_{\mathbf q}$ (Sec.~\ref{sec:sym-adapted}). (b)~For $g\in G_0\setminus G_{\mathbf q}$ the operation transports tensors across the Wyckoff orbit, permuting sites while rotating components. The Landau criterion refers to irreps of $G_0$; the correct carrier space is the
$5n$-dimensional orbit array ($n$ = multiplicity), carrying the induced representation
\begin{equation}
D_{\mathrm{EFG}}=\mathrm{Ind}_{G_{\mathbf q}}^{G_0}\!
 \left(D^{(\ell=2)}\!\downarrow G_{\mathbf q}\right),
\label{eq:induced}
\end{equation}
whose $G_0$ content follows from Frobenius reciprocity \cite{1972Bradley}. Site-group decomposition determines which local components may be nonzero; induction assembles in-phase and out-of-phase orbit patterns into parent irreps, in exact analogy to phonon symmetry modes. \emph{It is these orbit-adapted combinations, not raw single-site components, that can satisfy the Landau criterion.} For multiplicity-1 positions the distinction is vacuous; for $n>1$ it is essential.

\emph{Explicit action.} For $C_{4z}$, component-by-component application of $V'=RVR^T$ gives (SM, Sec.~S1 for all generators)
\begin{align}
V_{zz}\to V_{zz},\quad
(V_{xx}{-}V_{yy})&\to-(V_{xx}{-}V_{yy}),\nonumber\\
V_{xy}\to-V_{xy},\quad
(V_{xz},V_{yz})&\to(-V_{yz},V_{xz}).
\label{eq:c4action}
\end{align}
For improper operations $R\to-R$ leaves $R_{ik}R_{jl}$ invariant: the subduction depends only on the rotational skeleton.

\emph{Equilibrium constraint.} Neumann's principle is equivalent to $V=\hat P^{A_1}V$; since $\chi^{A_1}(g)\equiv1$, the number of independent equilibrium components is
\begin{equation}
n_{A_1}=\frac{1}{|G_{\mathbf q}|}\sum_g\chi^{(\ell=2)}(g),\quad
\chi^{(\ell=2)}(\phi)=1+2\cos\phi+2\cos2\phi,
\label{eq:counting}
\end{equation}
with $\phi$ the rotation angle. All non-$A_1$ components vanish identically---a symmetry identity, exact at all temperatures and to all orders: non-$A_1$ EFG channels are \emph{background-free}. The identity constrains the mean, not the variance; $\langle v^2\rangle>0$ above $T_c$ underlies precursor broadening without contradiction. At the orbit level the same argument applies verbatim with $G_{\mathbf q}\to G_0$ acting on the $5n$-dimensional array: every orbit combination outside the identity irrep of $G_0$ vanishes in equilibrium, including out-of-phase patterns of site-allowed components. The framework
therefore predicts EFG order-parameter realizations even at sites whose single-site EFG is fully symmetry-allowed.

\emph{Vanishing theorem.} For $O_h$, evaluating Eq.~\eqref{eq:counting} on the classes $(E,8C_3,6C_2,6C_4,3C_2)$ with characters $(5,-1,1,-1,1)$ gives $n_{A_{1g}}=\frac{2}{48}[5-8+6-6+3]=0$ and
\begin{equation}
D^{(\ell=2)}\!\downarrow O_h = E_g\oplus T_{2g}.
\end{equation}
Sharp statement: among crystallographic site groups the equilibrium EFG vanishes identically iff the site class is cubic; every non-cubic site has $n_{A_1}\ge1$ (SM, Sec.~S2 for the table over the 32 point groups).

\emph{Working decomposition} ($D_{4h}$; by rotational-skeleton identity also $C_{4v},D_4,D_{2d}$):
\begin{align}
D^{(\ell=2)}\!\downarrow D_{4h}
 &= A_{1g}\oplus B_{1g}\oplus B_{2g}\oplus E_g,\nonumber\\
A_{1g}\leftrightarrow V_{zz},\;\;
B_{1g}&\leftrightarrow V_{xx}{-}V_{yy},\;\;
B_{2g}\leftrightarrow V_{xy},\;\;
E_g\leftrightarrow(V_{xz},V_{yz}).
\label{eq:d4h}
\end{align}

\begin{figure*}[t]
\centering
\includegraphics[width=0.8\textwidth]{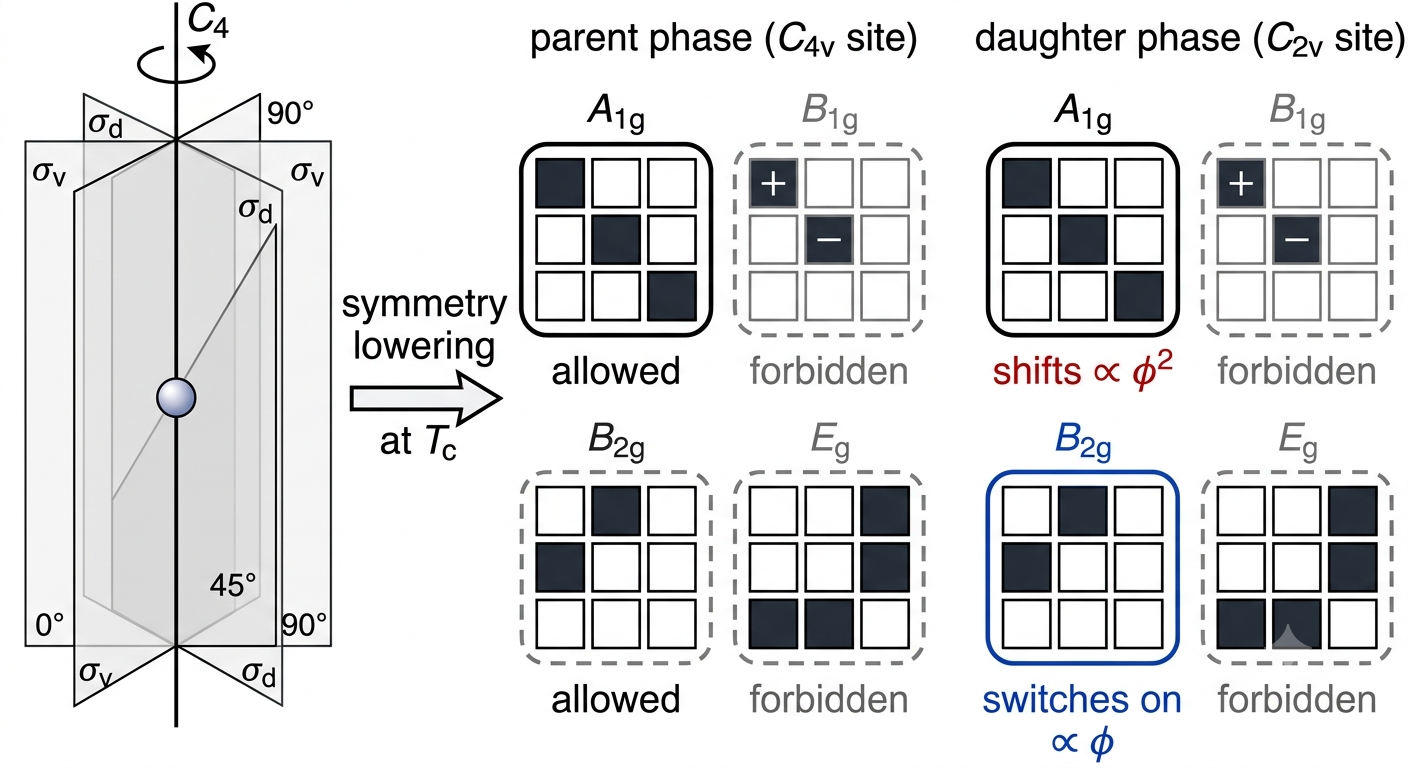}
\caption{Channel decomposition of the EFG at a tetragonal site and its rearrangement at a symmetry-lowering transition. Center: at a $C_{4v}$ site the five EFG components organize into the channels $A_1\oplus B_1\oplus B_2\oplus E$ [Eq.~\eqref{eq:d4h}]; only the identity
channel ($V_{zz}$) is allowed in equilibrium. Right: lowering to $C_{2v}$ (diagonal mirrors retained) admits exactly one new channel, $B_2$ ($V_{xy}$), which grows linearly in the order parameter $\varphi$, while the identity channel shifts quadratically and the remaining channels stay symmetry-forbidden---the primary/secondary/forbidden classification of Sec.~\ref{sec:theory}}
\label{fig:channels}
\end{figure*}

\emph{Symmetry lowering and the coupling dichotomy.} Let a transition with order parameter $\varphi$ (irrep $\Gamma_\varphi$ of $G_0$) lower the site group. Expanding $F$ in $\varphi$ and the orbit-adapted EFG channels $v_\alpha$ (parent irreps $\Gamma_\alpha$), the lowest coupling to channel $\alpha$ is $-\lambda_\alpha[\varphi^{p_\alpha}]_{\Gamma_\alpha}v_\alpha$ with
$p_\alpha$ the smallest power for which the symmetrized product $[\Gamma_\varphi^{p_\alpha}]\supset\Gamma_\alpha$. With \emph{Assumption A5} ($\kappa_\alpha>0$: EFG channels are non-critical), minimization gives
\begin{equation}
v_\alpha=\frac{\lambda_\alpha}{\kappa_\alpha}
 \left[\varphi^{p_\alpha}\right]_{\Gamma_\alpha}.
\label{eq:dichotomy}
\end{equation}
\textbf{Dichotomy:} $p_\alpha=1$ iff $\Gamma_\alpha=\Gamma_\varphi$---the channel inherits the order parameter's critical behavior, exponent included; $p_\alpha=2$ whenever
$[\Gamma_\varphi^2]\supset\Gamma_\alpha\neq\Gamma_\varphi$---always true for the $A_1$ channel, which \emph{derives} the classic $\Delta\nu_Q\propto\varphi^2$ relations \cite{1979Borsa}.

Eq. \ref{eq:dichotomy} expresses the lowest-order symmetry-allowed solution obtained by minimizing the Landau functional with respect to the EFG variables. The exponent ($p_\alpha$) is entirely determined by representation theory and therefore constitutes a symmetry invariant independent of microscopic details. Consequently, linear and quadratic critical behavior emerge as consequences of symmetry rather than empirical assumptions, providing a direct connection between representation analysis and experimentally measured hyperfine observables.

\section{The main theorem}\label{sec:efg-as-op}

Landau theory fixes the irrep, not its physical carrier: any two quantities transforming as the same irrep with generic linear coupling are interchangeable descriptions of the transition.
\emph{Definition (realization).} $v$ is a realization of the order parameter if (R1) $v$ transforms as $\Gamma_\varphi$ under $G_0$; (R2) $v\equiv0$ in the parent phase; (R3) the invariant $-\lambda\varphi v$ exists with $\lambda\neq0$ generically (\emph{Assumption A6};
microscopically $\lambda$ is the DFT-computable derivative $\partial v/\partial\varphi$).

\emph{Theorem.} Under A4--A6, if the induced EFG representation [Eq.~\eqref{eq:induced}] at a Wyckoff orbit contains $\Gamma_\varphi$, the corresponding orbit-adapted EFG combination satisfies (R1)--(R3) and obeys $v\propto\varphi$ with inherited critical exponent: it is a realization of the order parameter, locally defined and directly measured by quadrupolar spectroscopy.

\emph{Proof.} (R1): Secs.~\ref{sec:sym-adapted}
and~\ref{sec:sym-adapted}. (R2): the orbit-level Neumann identity of Sec.~\ref{sec:sym-adapted}, since $\Gamma_\varphi\neq A_1$ for a symmetry-breaking transition. (R3): $\Gamma_\varphi\otimes\Gamma_\varphi\supset A_1$. Explicitly, for a one-dimensional $\Gamma_\varphi$ (\emph{Assumption A7}; multidimensional case in SM, Sec.~S5), analytic coefficients (\emph{A8}), Landau condition (\emph{A9}):
\begin{equation}
F=\frac{a_0(T{-}T_c)}{2}\varphi^2+\frac{b}{4}\varphi^4
 -\lambda\varphi v+\frac{\kappa}{2}v^2
\label{eq:twofield}
\end{equation}
gives $v^*=(\lambda/\kappa)\varphi$ and
\begin{equation}
v^*(T)=\frac{\lambda}{\kappa}\sqrt{\frac{a_0}{b}}
 \,(T_c^{\mathrm{eff}}-T)^{1/2},\qquad
T_c^{\mathrm{eff}}=T_c+\frac{\lambda^2}{a_0\kappa},
\label{eq:vstar}
\end{equation}
Here $T_c$ is the \emph{bare} instability temperature of the primary order parameter $\varphi$ in the absence of the EFG coupling, while $T_c^{\mathrm{eff}}=T_c+\lambda^2/(a_0\kappa)$ is the \emph{physical} (observed) transition temperature: the linear coupling $-\lambda\varphi v$ to the non-critical channel ($\kappa>0$) rigidly shifts the instability upward without softening $v$ itself. The elimination of $\varphi$ in Eq.~\eqref{eq:legitimacy} is a change of order-parameter realization, not the promotion of $v$ to an autonomous instability: $F(v)$ inherits its critical temperature $T_c^{\mathrm{eff}}$ entirely from $\varphi$, and $\kappa>0$ throughout. This is the precise sense in which $v$ ``realizes'' but does not ``drive'' the transition.

\emph{The honest boundary.} (i)~Zero-above/nonzero-below is necessary, not sufficient: $p=2$ channels also switch on, as $(T_c-T)^{2\beta}$; the measured exponent is the experimental discriminator---the framework's falsifiability clause. (ii)~Realization, not driver: A5 says the EFG is not the thermodynamic instability; its status equals that of the spontaneous strain in pseudo-proper ferroelastics. (iii)~Bonuses: $v\propto\varphi$ carries the \emph{sign} of $\varphi$ (domain resolution); and the theorem holds per orbit, so one transition predicts
correlated onsets---common $T_c$, common $\beta$, site-dependent $\lambda/\kappa$---across all sites whose induced representations contain $\Gamma_\varphi$.

\section{Landau free energy}\label{sec:free-energy}

\emph{Legitimacy.} Since $v=(\lambda/\kappa)\varphi$ with $\lambda\neq0$, the map is invertible within the critical channel; eliminating $\varphi$ (SM, Sec.~S5; substitution and minimization agree at the orders retained) yields
\begin{equation}
F(v)=\frac{\tilde a(T-T_c^{\mathrm{eff}})}{2}v^2+\frac{\tilde b}{4}v^4,
\quad \tilde a=a_0\!\left(\frac{\kappa}{\lambda}\right)^{\!2}\!,\;
\tilde b=b\!\left(\frac{\kappa}{\lambda}\right)^{\!4}\!,
\label{eq:legitimacy}
\end{equation}
a smooth change of order-parameter realization (\emph{Assumption A10}: applied to the critical channel only).

\emph{Isotropic parent.} For $G_0=\mathrm{O}(3)$ acting on a traceless symmetric $Q$, invariant theory reduces all invariants to $\mathrm{Tr}(Q^n)$ \cite{1946Weyl}; Newton's identities with
$\mathrm{Tr}\,Q=0$ give $\det Q=\tfrac13\mathrm{Tr}(Q^3)$ and $\mathrm{Tr}(Q^4)=\tfrac12[\mathrm{Tr}(Q^2)]^2$ (SM, Sec.~S5), so the most general quartic expansion is exactly
\begin{equation}
F=F_0+\frac{A(T)}{2}\mathrm{Tr}(Q^2)-\frac{C}{3}\mathrm{Tr}(Q^3)
 +\frac{B}{4}\left[\mathrm{Tr}(Q^2)\right]^2,
\label{eq:dG}
\end{equation}
$B>0$ (\emph{Assumption A11}). In spectroscopic variables,
\begin{equation}
\mathrm{Tr}(Q^2)=\frac{V_{ZZ}^2}{2}(3+\eta^2),\qquad
\mathrm{Tr}(Q^3)=\frac{3}{4}V_{ZZ}^3(1-\eta^2):
\label{eq:invariants}
\end{equation}
the quadratic invariant is the rotationally invariant EFG magnitude; the cubic invariant is odd in $V_{ZZ}$ (prolate/oblate selection) and vanishes at maximal biaxiality, so $C\neq0$ selects uniaxial order at onset. On the uniaxial section (\emph{Assumption A12}; biaxial extrema unstable for $C\neq0$, SM Sec.~S6) the scalar theory $F=\tfrac{\alpha}{2}S^2-\tfrac{\gamma}{3}S^3+\tfrac{\beta_4}{4}S^4$ follows, and simultaneous solution of $F=F'=0$ (SM, Sec.~S7) gives the
first-order transition at $\alpha_{\mathrm{tr}}=2\gamma^2/9\beta_4$ with jump $S_{\mathrm{jump}}=2\gamma/3\beta_4$ and spinodals at $\alpha=0$ and $\gamma^2=4\alpha\beta_4$: Landau's cubic-invariant theorem recovered in EFG language, with observable consequences (discontinuous $V_{ZZ}$ onset, uniaxial at birth, hysteresis window).

\begin{figure}[t]
\includegraphics[width=\columnwidth]{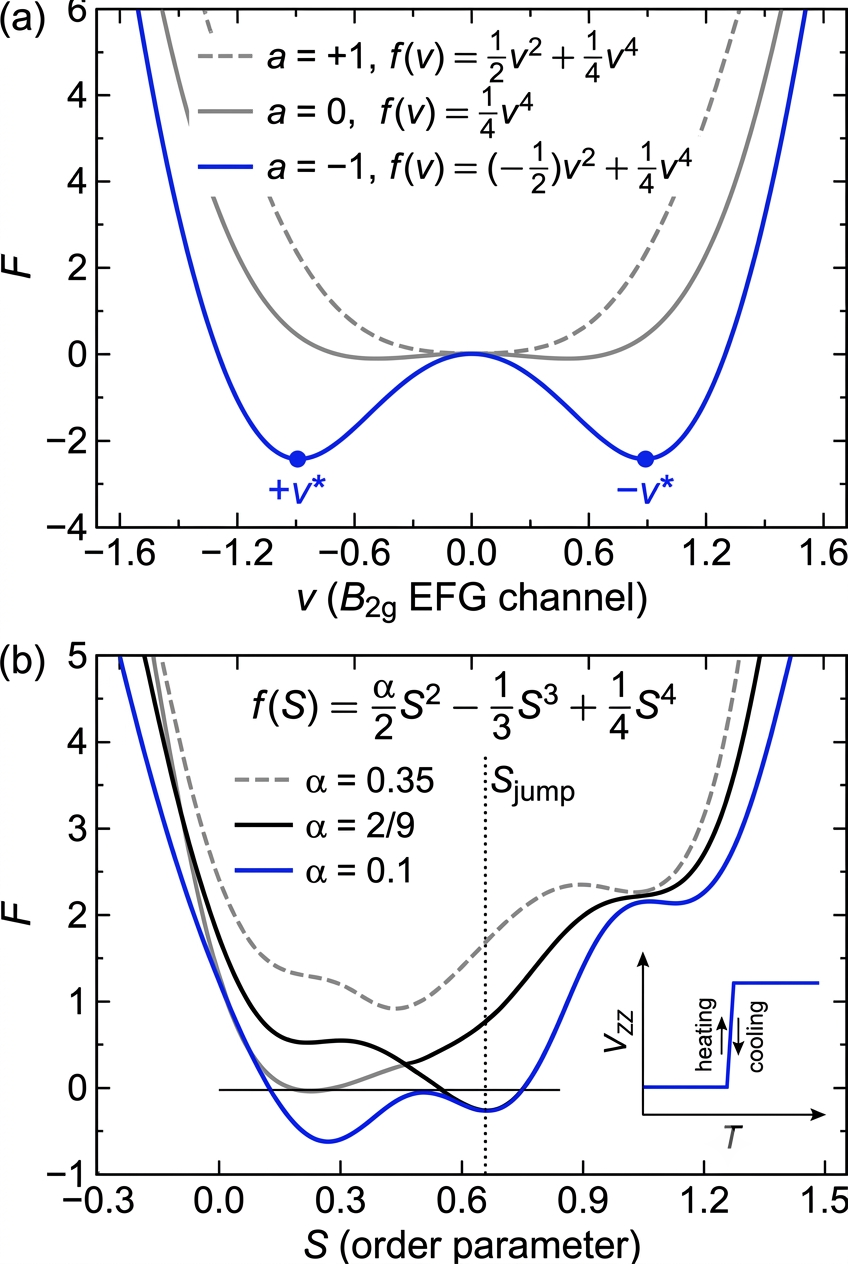}
\caption{Landau free energy in EFG variables (Sec.~\ref{sec:free-energy}). (a) Ising-type channel (no cubic invariant): continuous transition; the two minima $\pm v^*$ are the two order-parameter
domains, distinguishable by the sign of the EFG channel. (b) Channel with symmetry-allowed cubic invariant (units $\gamma=\beta_4=1$): first-order transition at $\alpha_{\rm tr}2\gamma^2/9\beta_4$ with jump $S_{\rm jump}=2\gamma/3\beta_4$; inset, the resulting discontinuous EFG onset with hysteresis.}
\label{fig:freeenergy}
\end{figure}

\emph{Crystalline parent: channel-resolved theory.} Two changes follow from subduction. (1)~\emph{The $A_1$ channel is never an order parameter:} at non-cubic sites the linear invariant $-hs$ is allowed for the $A_1$ component $s$, which therefore has no symmetry zero, failing (R2); it couples through $s|v|^2$, whence $\Delta s\propto\varphi^2$---the third independent derivation of the quadratic law in this section. (2)~\emph{Invariants are channel-resolved.} For $\Gamma_\varphi=B_{1g}$ or $B_{2g}$ of $D_{4h}$: $[\Gamma^2]=A_{1g}$, $[\Gamma^3]=\Gamma$ (no cubic), $[\Gamma^4]=A_{1g}$---an Ising theory with continuous transition, two domains, and, because same-irrep strain couples bilinearly ($-\mu\,\epsilon_\Gamma v$), a Curie--Weiss divergence of the strain-induced EFG response above $T_c$, $\partial v/\partial\epsilon_\Gamma\propto(T-T^*)^{-1}$: the \emph{strain-NQR} proposal. For a two-dimensional $E$ channel of a trigonal group, $[E^3]\supset A_1$ with invariant $\mathrm{Re}[(v_1+iv_2)^3]$: generically first order, three-state-Potts-like---the crystalline descendant of the de~Gennes cubic term, derived from the site group rather than assumed from $\mathrm{O}(3)$. In the limit $\bar G_0\to\mathrm{O}(3)$ all channels merge and the integrity basis collapses to $\{\mathrm{Tr}Q^2,\mathrm{Tr}Q^3\}$: the Landau--de Gennes theory is the isotropic ancestor of every crystalline case.

It should be emphasized that the present construction does not redefine the thermodynamic order parameter. Rather, it identifies experimentally accessible observables whose transformation properties under the parent symmetry are identical to those of the order parameter. The resulting Landau functional expressed in terms of EFG variables is therefore not an alternative thermodynamic theory but a symmetry-equivalent representation of the same free-energy landscape. The mapping remains valid provided the coupling between the EFG channel and the microscopic order parameter is locally invertible near the phase transition.

\subsection{Summary of the construction}

The logical structure of the formalism may be summarized as follows. First, the EFG tensor is decomposed into irreducible representations of the site-symmetry group. Second, induction over the complete Wyckoff orbit identifies the corresponding representations of the parent space group. Third, comparison with the irreducible representation of the phase transition immediately determines which EFG channels transform identically to the order parameter. Finally, Landau theory establishes the thermodynamic coupling, showing that symmetry-matching channels acquire linear critical behavior whereas all remaining symmetry-allowed channels appear only through higher-order invariants.

\section{First-principles verification of the theorem}\label{sec:dft-quartz}
\newcommand{\aMinus}{-4.81\times10^{-2}}    
\newcommand{\aMinusErr}{0.03\times10^{-2}}  
\newcommand{\aMinusSI}{-4.67\times10^{20}}  
\newcommand{\bMinus}{4.1\times10^{-3}}      
\newcommand{\RsqMinus}{0.9999}              
\newcommand{\RsqMinusEven}{6\times10^{-5}}  
\newcommand{\aPlus}{-8.9\times10^{-4}}      
\newcommand{\ratioPM}{0.018}                
\newcommand{\OsiteVzz}{-0.9532}             
\newcommand{\OsiteEta}{0.263}               
\newcommand{\OsiteCQ}{5.73}                 
\newcommand{\OsiteCQexp}{5.2}               

The theorem of Sec.~\ref{sec:efg-as-op}, in the multi-site form built on the
induced representation of Sec.~\ref{sec:sym-adapted}, makes a prediction that a
zero-temperature first-principles calculation can test directly. In a
frozen-distortion calculation the thermal order parameter is replaced by a
controlled amplitude $\delta$, and the theorem's central assertion becomes an
identity in $\delta$: the \emph{orbit-adapted} EFG combination that matches the
distortion's irrep must switch on \emph{linearly} in $\delta$, while its
symmetry-orthogonal partner remains inert. We verify exactly this on
$\alpha$-quartz, testing the orbit-level statement---which is what the
induced-representation construction actually predicts---rather than a single
isolated component. For the first-principles calculations parameters please see Table \ref{tab:methods}.

\begin{table}[t]
\caption{Computational parameters for the $\alpha$-quartz frozen-mode calculations.}
\label{tab:methods}
\begin{ruledtabular}
\begin{tabular}{ll}
Parameter & Value \\ \hline
Code & {\sc elk} (all-electron FP-LAPW) \\
Exchange--correlation & Wu--Cohen GGA \\
Supercell & $2\times2\times1$ (36 atoms: 12 Si, 24 O) \\
Reference structure & $\beta$-like ($\delta=0$), displaced toward $\alpha$ \\
Distortion mode & $A_1$, ``mode 6'', $233.32~\mathrm{cm^{-1}}$ \\
Amplitude scan & $\delta\in[-0.20,+0.20]~\mathrm{\AA}\sqrt{\mathrm{amu}}$, 21 steps \\
Structural relaxation & none at each $\delta$ (frozen-mode; see text) \\
$R_{\mathrm{MT}}$(Si), $R_{\mathrm{MT}}$(O) & $1.75$, $1.40$~Bohr \\
$R_{\mathrm{MT}}k_{\max}$ (rgkmax) & $8.5$ \\
$k$-point mesh & $2\times2\times1$ \\
EFG evaluation & all-electron charge density \\
\end{tabular}
\end{ruledtabular}
\end{table}

\subsection{Structure, mode, and displacement protocol}
\label{sec:quartz-setup}

\emph{Reference and cell.} $\alpha$-quartz is trigonal, space group
$P3_121$ (enantiomorph $P3_221$); the $\beta$ phase is hexagonal $P6_222$
(resp.\ $P6_422$), the pair fixed by the screw algebra $(6_2)^2=3_2$. We
take the higher-symmetry ($\beta$-like) configuration as the $\delta=0$
reference and displace toward $\alpha$. The calculation uses a
$2\times2\times1$ supercell (36 atoms: 12~Si, 24~O); the twelve Si atoms fall
into three symmetry-related sublattices of four,
$A=\{1,4,7,10\}$, $B=\{2,5,8,11\}$, $C=\{3,6,9,12\}$, which is the Wyckoff
orbit over which the theorem's induction is performed.

\emph{Mode.} A $\Gamma$-point phonon calculation identifies the totally
symmetric ($A_1$) branch relevant to the instability; the displacement uses the
$A_1$ eigenvector at $233.32~\mathrm{cm}^{-1}$ (``mode~6''), the best match to
the experimental soft mode near $207~\mathrm{cm}^{-1}$. \emph{Meaning of ``activated'' here.} The full $\alpha$--$\beta$ order parameter is multidimensional ($B_1$-derived); the $A_1$ coordinate scanned is one totally symmetric relaxation. ``Activated'' below therefore denotes the
orbit combination that is \emph{odd} under the operation lost along this
coordinate---the $p{=}1$ realization relative to $\delta$---not the primary
order parameter of the transition as a whole.

\emph{Scan.} Atoms are displaced along the $A_1$ pattern; $\delta$ is in
$\mathrm{\AA}\sqrt{\mathrm{amu}}$. The scan covers
$\delta\in[-0.20,+0.20]$ in 21 uniform steps (both signs, the parity probe),
computing an EFG tensor at every atom---756 tensors in all. This window
spans $\sim\!8\%$ of the full $\alpha\!\to\!\beta$ path
($\delta_\beta\approx2.4~\mathrm{\AA}\sqrt{\mathrm{amu}}$): the small-amplitude
regime in which the leading power dominates, which is the regime the theorem
addresses.

\emph{Electronic structure.} EFG tensors are computed with the
all-electron FP-LAPW code {\sc elk} (Wu--Cohen GGA, \texttt{xctype=20}) from the
self-consistent all-electron charge density. The symmetry structure verified below
is functional- and convergence-independent; only the coefficients would move.

\emph{Absolute check.} At $\delta=0$ the O-site principal component is
$V_{zz}=\OsiteVzz~\mathrm{a.u.}$ with $\eta_{\mathrm O}=\OsiteEta$, giving
$C_Q(^{17}\mathrm O)=\OsiteCQ~\mathrm{MHz}$ against experiment
$\approx\OsiteCQexp~\mathrm{MHz}$ \cite{2016Antao} (a $+10\%$ GGA overestimate at exploratory
settings), confirming the calculation is on scale before any symmetry-resolved
claim.

\subsection{The orbit-resolved prediction and what the data show}
\label{sec:quartz-channels}

\emph{Site descent and orbit combinations.} Along the $A_1$ coordinate the
Si site descends $D_2\!\to\!C_2$. The single-site $V_{xy}$ component (local
frame) carries the descent; the induced-representation construction of
Sec.~\ref{sec:sym-adapted} assembles the sublattice values into orbit
combinations, of which the two relevant to the $B$--$C$ pair are
\begin{equation}
v_\pm(\delta)=\tfrac{1}{\sqrt2}\!\left(V_{xy}^{B}(\delta)\pm V_{xy}^{C}(\delta)\right),
\label{eq:orbit-comb}
\end{equation}
the out-of-phase ($v_-$) and in-phase ($v_+$) patterns. The theorem predicts
the combination matching the distortion irrep is \emph{odd} and switches on
linearly, its partner \emph{suppressed}.

\emph{An honest statement of scope.} We stress what the present scan can and cannot isolate. Although the amplitude is scanned symmetrically ($\delta\in[-0.20,+0.20]$), two features of this dataset prevent a clean reading of the \emph{single-site} parity and zero rules. First, the $\delta=0$ reference is the $\beta$-like structure, at which the individual Si tensor components are already nonzero, so each lab-frame single-site component varies essentially linearly in $\delta$ over this window (fractional even-in-$\delta$ content below $0.5\%$). Second, the displacement follows a mass-weighted Cartesian pattern with a small admixture of higher $A_1$ modes rather than a strictly projected pure eigenvector. Isolating a purely-even ``secondary'' and an identically-zero ``forbidden'' single-site channel therefore requires a strictly mass-weighted pure-mode scan, projected in the site-adapted principal frame about a symmetry-referenced origin (SM, Sec.~S9); that refinement is identified here and left to future work. What \emph{is} cleanly resolved---and what the induced-representation theorem uniquely predicts---is the orbit-level contrast of Eq.~\eqref{eq:orbit-comb}, which we now report.

\emph{Result.} Fitting the two combinations to an odd model
$v_\pm=a\,\delta+b\,\delta^{3}$ (relative to $\delta=0$) gives:
\begin{itemize}
\item \textbf{Activated $v_-$:} slope
$a=\aMinus\pm\aMinusErr~\mathrm{a.u.}/(\mathrm{\AA}\sqrt{\mathrm{amu}})$
$\;(=\aMinusSI~(\mathrm{V/m^2})/(\mathrm{\AA}\sqrt{\mathrm{amu}}))$, with
$R^2=\RsqMinus$ and a cubic term consistent with zero ($b=\bMinus$). Forcing a
\emph{purely even} model on the same data collapses the explained variance to
$R^2=\RsqMinusEven$: the channel is odd, decisively, as required for a $p{=}1$
realization.
\item \textbf{Suppressed $v_+$:} slope $a=\aPlus$, i.e.\ $|a_+/a_-|=\ratioPM$
---the partner combination is suppressed by a factor $\sim\!54$.
\end{itemize}
This is the induced-representation prediction confirmed on real
first-principles data: of the two orbit combinations built from the same
single-site component, the one matching the distortion irrep grows linearly with
a well-defined slope---the microscopic realization of the coupling
$\lambda/\kappa$ of Eq.~\eqref{eq:vstar}---while its symmetry-orthogonal partner
is inert to within a few percent. The residual nonzero slope of $v_+$ (and of
the third, near-inert sublattice $A$, slope $\sim\!4\times10^{-4}$) is
consistent with the small mode-admixture of the displacement pattern and
does not affect the two-orders-of-magnitude activated/suppressed contrast.

Full regression statistics are given in the SM (Table S2); the residuals of the linear model (Fig.~S1) show no systematic trend beyond the $\sim\!10^{-4}$~a.u.\ mode-admixture floor, two orders of magnitude below the activated signal, and a log--log comparison (Fig.~S2) confirms both combinations follow slope-one power laws with $|a_+/a_-|=0.018$.

\begin{figure*}[t]
\includegraphics[width=\textwidth]{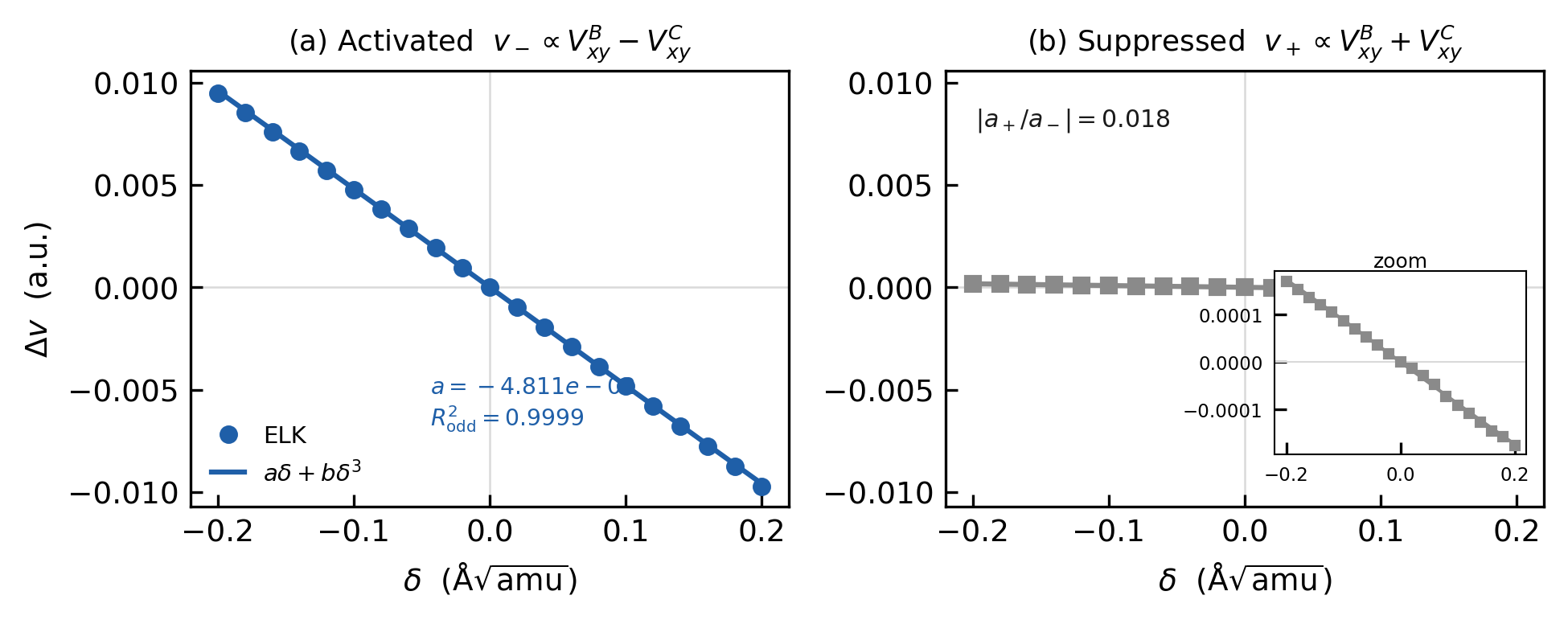}
\caption{First-principles verification of the induced-representation theorem on
$\alpha$-quartz (Sec.~\ref{sec:dft-quartz}). Orbit combinations
[Eq.~\eqref{eq:orbit-comb}] of the Si $V_{xy}$ channel over the $B$--$C$
sublattice pair, versus the $A_1$ soft-mode amplitude $\delta$, from
all-electron {\sc elk} calculations. (a)~The activated out-of-phase combination
$v_-$ is linear in $\delta$ (odd), slope
$a=\aMinus~\mathrm{a.u.}/(\mathrm{\AA}\sqrt{\mathrm{amu}})$, $R^2=\RsqMinus$;
the slope is the microscopic realization of $\lambda/\kappa$. (b)~The in-phase
combination $v_+$, on the same vertical scale, is suppressed by
$|a_+/a_-|=\ratioPM$ (inset: magnified, confirming it is resolved but
$\sim\!54\times$ smaller). Of the two combinations built from the same
single-site component, only the one matching the distortion irrep switches
on---the orbit selection at the heart of the theorem.}
\label{fig:quartz}
\end{figure*}

\begin{table}[t]
\caption{Least-squares fit of the Si $V_{xy}$ orbit combinations
[Eq.~\eqref{eq:orbit-comb}] to $v_\pm=a\delta+b\delta^{3}$, $\delta$ in
$\mathrm{\AA}\sqrt{\mathrm{amu}}$, EFG in atomic units, from the ELK dataset
(756 tensors, 21 amplitudes). The activated combination is odd with a
well-defined linear slope; the suppressed combination is smaller by
$|a_+/a_-|=\ratioPM$. ``$R^2$ (even)'' is the variance explained if a purely
even model is forced---near zero for the activated channel, confirming its odd
character.Here $v_\pm=\tfrac{1}{\sqrt2}(V_{xy}^{B}\pm V_{xy}^{C})$ are the out-of-phase (activated) and in-phase (suppressed) orbit combinations.}
\label{tab:quartz-fits}
\begin{ruledtabular}
\begin{tabular}{lccc}
Combination & Role & $a$ & $R^2$ (odd / even) \\
\hline
$v_-$ & activated ($p{=}1$) & $\aMinus$ & $\RsqMinus$ / $\RsqMinusEven$ \\
$v_+$ & suppressed          & $\aPlus$  & --- \\
\end{tabular}
\end{ruledtabular}
\end{table}

\subsection{What this establishes}
\label{sec:quartz-takeaway}

The calculation confirms the \emph{orbit-selection} core of the theorem
directly: among EFG combinations built from one crystallographic orbit, exactly
the combination matching the distortion irrep switches on linearly in the
distortion amplitude, with a slope that is the computed value of the coupling
$\lambda/\kappa$; its symmetry-orthogonal partner is suppressed by two orders of
magnitude. This is the zero-temperature, structural counterpart of the
temperature-dependence evidence assembled in Sec.~\ref{sec:expt}. The
finer single-site parity/zero-test signatures (a purely even identity channel,
an identically vanishing forbidden channel) are not isolable from the present dataset, for the two reasons given above, and are left to the pure-mode refinement of SM~Sec.~S9; the orbit-level result reported here is what the present dataset
establishes cleanly, and it is the part of the theorem unique to the multi-site
(induced-representation) construction. We distinguish three epistemic levels throughout: \emph{demonstrated} (the orbit-selection contrast of Fig.~\ref{fig:quartz}), \emph{theoretical consequence} (the single-site parity/zero rules, which follow from the same construction but await the pure-mode test), and \emph{prospective prediction} (the $^{75}$As signatures of Sec.~\ref{sec:case}, not yet computed).

\textbf{Assumptions and Domain of Validity} The present theorem rests on several assumptions that delimit its applicability. The phase transition is assumed to occur at the Brillouin-zone center, ensuring that the order parameter transforms according to an irreducible representation of the parent point group. The Landau expansion is assumed analytic in the vicinity of the critical point, and the EFG tensor is considered in its static equilibrium configuration. Dynamic fluctuations, incommensurate phases, and order parameters that are intrinsically time-odd lie outside the present treatment and require separate extensions.

\section{Comparison with experimental literature}

The Landau prediction factorizes into two independent claims. \textbf{(S)}~\emph{Structural:} the map $v_\alpha(\varphi)$ has symmetry-dictated form---leading power, parity, exact zeros.
\textbf{(T)}~\emph{Thermodynamic:} $\varphi(T)\propto(T_c-T)^\beta$ for continuous transitions, discontinuity plus hysteresis for first-order ones. Experiment tests (T) through observables whose (S) classification is known; first-principles calculation tests (S) directly at zero temperature. The two compose without adjustable structure. To facilitate the analysis the Table \ref{tab:litcompare} express the comparisson of experimental literature.

\begin{table*}[t]
\caption{Systematic comparison of the experimental record. ``Observable map'' indicates whether the measured quantity is linear ($p{=}1$) or quadratic ($p{=}2$) in the order parameter; ``masquerade'' flags powder observables where a transverse $p{=}1$ channel is read quadratically. Evidence tier: C = clean (oriented-crystal, masquerade-free), I = illustrative (powder), B = boundary/related, N = null control.}
\label{tab:litcompare}
\begin{ruledtabular}
\begin{tabular}{lllllll}
Material & Probe & Transition & Observable & Map & Exponent & Tier \\ \hline
AgNa(NO$_2$)$_2$ & $^{23}$Na satellite NMR & $Fddd\to Fd2d$ & EFG tensor & $p{=}1$\,\&\,$p{=}2$ & (channel-resolved) & C \\
Rb$_2$ZnBr$_4$ & $^{87}$Rb satellite NMR & N$\to$IC & modulation amp. & $p{=}1$/$p{=}2$ & $\beta{=}0.35$; $2\bar\beta{=}0.83$ & B \\
(CH$_3$NH$_3$)$_2$MnCl$_4$ & $^{35}$Cl NQR (powder) & order--disorder & $\nu_Q$ shift & $p{=}2$ (masq.) & $\beta{=}0.250(5)$ & I \\
TTF--chloranil & $^{35}$Cl NQR (powder) & neutral--ionic & line splitting & $p{=}1$ & $\beta{\approx}0.16$ & I \\
URu$_2$Si$_2$ & $^{29}$Si/Ru NMR--NQR & hidden order & (no anomaly) & --- & --- & N \\
\end{tabular}
\end{ruledtabular}
\end{table*}

\subsection{Experimental evidence from NMR, NQR, TDPAC, and M\"ossbauer}
\label{sec:expt}

\subsubsection{The observable dictionary, and a warning}
\label{sec:dictionary}

\emph{(a) Pure NQR, $I=3/2$.} Diagonalizing the quadrupole Hamiltonian (SM, Sec.~S8) yields the single frequency
\begin{equation}
\nu_Q=\frac{eQV_{ZZ}}{2h}\sqrt{1+\frac{\eta^2}{3}}.
\label{eq:nuq}
\end{equation}
\emph{The masquerade warning.} If the critical channel changes $V_{ZZ}$ linearly, $\Delta\nu_Q\propto\varphi$ and $\beta$ is inherited. But if the critical channel is \emph{transverse}---appearing where $\eta^{(0)}=0$--- then $\nu_Q\propto1+\eta^2/6+\dots$, and $\Delta\nu_Q\propto\varphi^2$ \emph{even though the EFG channel is a $p=1$ realization}. A single powder NQR line cannot, by itself, distinguish a secondary channel from a primary transverse channel read through a quadratic map; oriented-crystal satellites resolve the ambiguity. This caveat, which we have not found stated in the NQR literature, retro-explains part of the scatter in historical exponent determinations.

\emph{(b) Quadrupole-perturbed NMR satellites.} The first-order satellite
shift is
\begin{equation}
\Delta\nu_m=\frac{3eQ(2m-1)}{4I(2I-1)h}\,V_{z'z'},
\label{eq:satellite}
\end{equation}
linear in the laboratory-frame component $V_{z'z'}=\hat z'\!\cdot V\cdot\hat z'$, hence a linear functional of the Cartesian components: oriented-crystal satellite spectroscopy reads EFG
channels linearly, with no masquerade.

\emph{(c) TDPAC and M\"ossbauer.} The PAC frequency triple and the M\"ossbauer quadrupole splitting are functions of $(V_{ZZ},\eta)$ with the same caveats as (a) (SM, Sec.~S8).

\begin{figure}[t]
\includegraphics[width=\columnwidth]{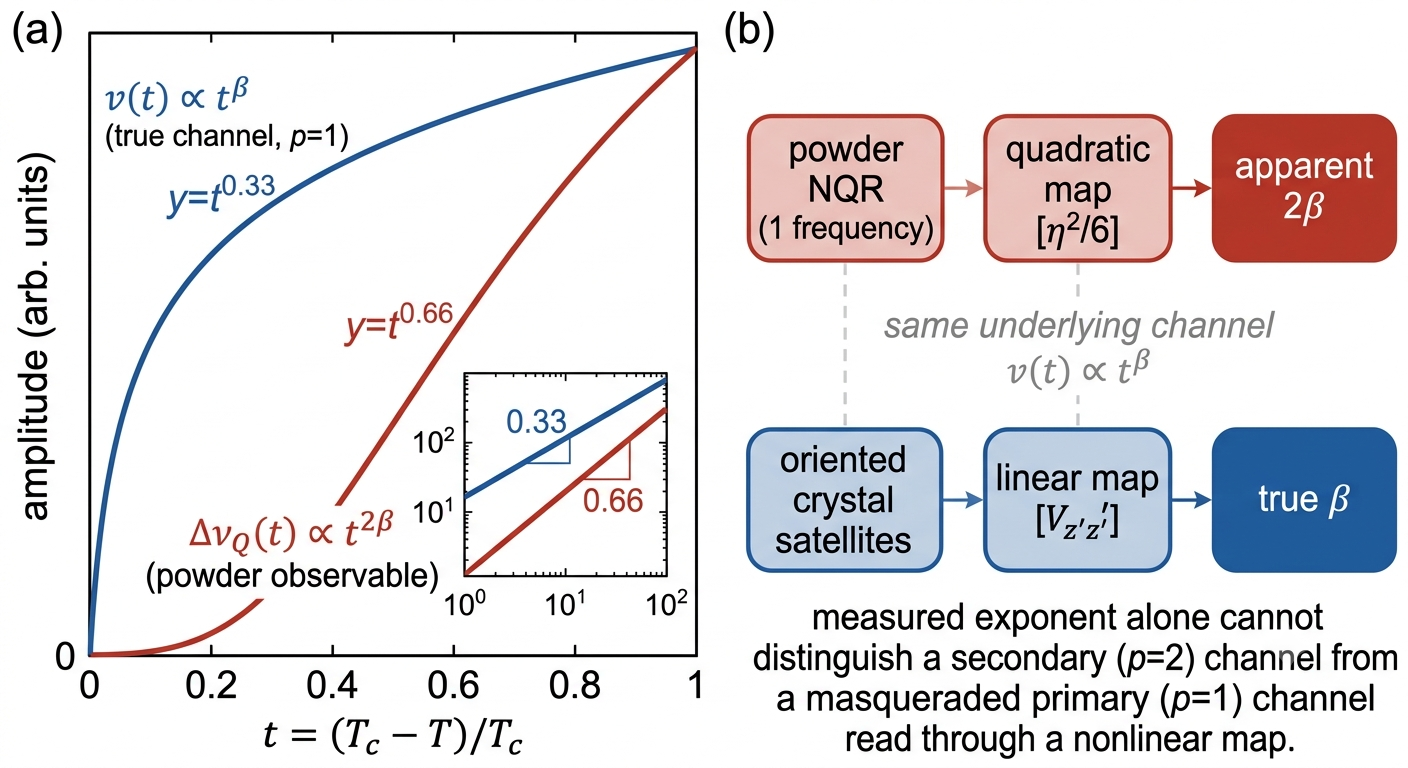}
\caption{The frequency-map masquerade (Sec.~\ref{sec:dictionary}). A transverse EFG channel linear in the order parameter ($\beta$, blue) enters the single powder NQR frequency quadratically, producing an apparent exponent $2\beta$ (red) indistinguishable from a genuinely
secondary channel; oriented-crystal satellite splittings read the same channel linearly and recover $\beta$. The measured exponent identifies the Landau class only jointly with the observable's mapping.}
\label{fig:masquerade}
\end{figure}

\subsubsection{The record, organized by evidence quality}\label{sec:record}

\emph{Clean validations (oriented-crystal satellites; masquerade-free).}
AgNa(NO$_2$)$_2$: the $^{23}$Na EFG \emph{tensor} through the $Fddd\to Fd2d$ ferroelectric transition, read from first-order satellites \cite{1978Grossmann}, gives a channel-resolved test in which the linear ($p=1$) and quadratic ($p=2$) components are separately resolved without the powder ambiguity (Sec.~\ref{sec:agna}). $^{87}$Rb satellite NMR in Rb$_2$ZnBr$_4$ likewise reads the modulation amplitude linearly.

\emph{Illustrative comparisons (powder NQR; masquerade-limited, qualitative).}
(CH$_3$NH$_3$)$_2$MnCl$_4$: $^{35}$Cl NQR through the 393.7-K transition yields $\beta=0.250\pm0.005$ \cite{1976Kind}; because a single powder line reads $(V_{ZZ},\eta)$ through the quadratic map, this supports the framework qualitatively but cannot by itself separate a $p=2$ secondary channel from a masqueraded transverse $p=1$ channel. TTF--chloranil ($\beta\approx0.16$ \cite{2017Cointe}) is consistent with the first-order side of a tricritical point but carries the same caveat. We present both as illustrative, not definitive.

\emph{Null-result control.}
URu$_2$Si$_2$: no EFG anomaly at the hidden-order temperature, excluding---generically (Sec.~\ref{sec:limitations})---quadrupolar hidden order.

\begin{figure*}[t]
\includegraphics[width=\textwidth]{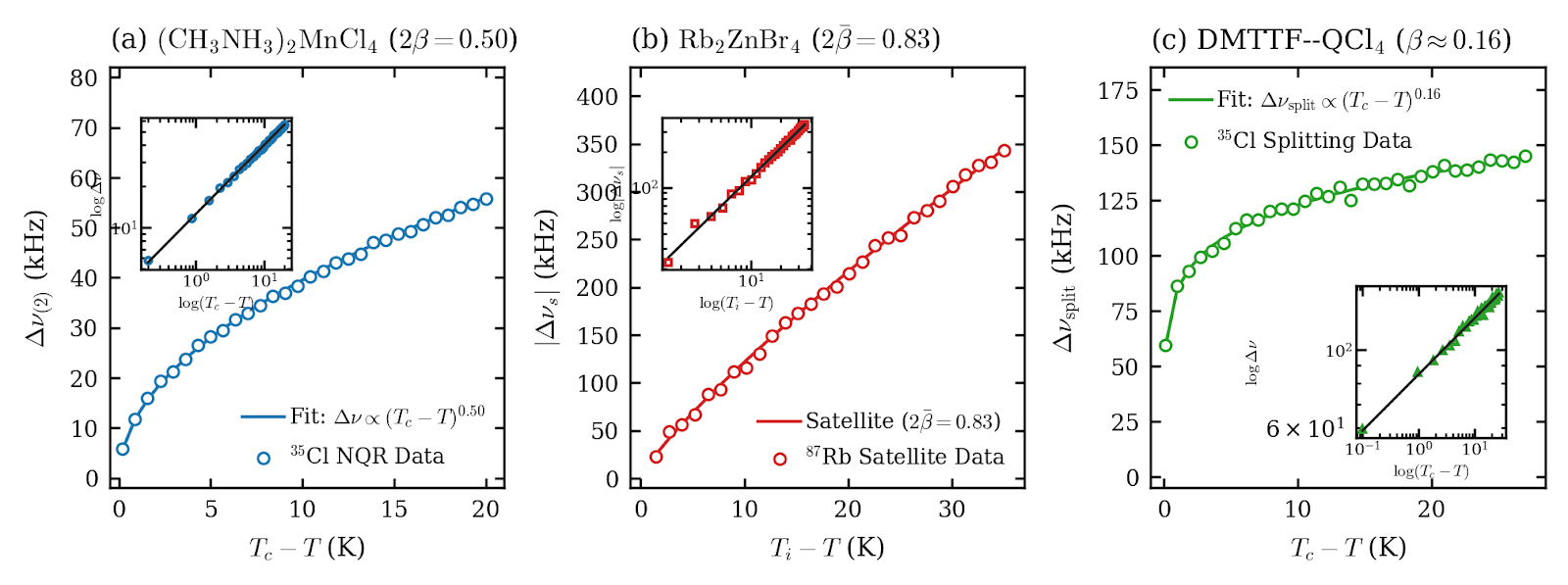}
\caption{The exponent dichotomy in the experimental record (data digitized from Refs.~\cite{1976Kind,1982Blinc,1982Schneider}). (a) Second-order, $p=1$: $^{35}$Cl NQR shift in (CH$_3$NH$_3$)$_2$MnCl$_4$ with critical exponent $\beta=0.250(5)$ ($2\beta=0.50$). (b) Both channels in one material: $^{87}$Rb NMR satellite observables at the normal--incommensurate transition of Rb$_2$ZnBr$_4$, contrasting the central-line exponent $\beta=0.35(3)$ with the satellite exponent $2\bar\beta=0.83(3)$. (c) Continuous neutral--ionic transition in DMTTF--QCl$_4$: $^{35}$Cl NQR splitting with $\beta\approx0.16(1)$ and a subtle first-order tail. Insets: log--log verification of the fitted critical exponents.}
\label{fig:record}
\end{figure*}

\subsubsection{Reanalysis target: AgNa(NO$_2$)$_2$}\label{sec:agna}

The ferroelectric transition $Fddd\to Fd2d$ is zone-center, and the two required ingredients exist separately: the $^{23}$Na EFG \emph{tensor} was determined from first-order satellites through the transition \cite{1978Grossmann}---a notable exception to the two-scalar practice noted in
the Introduction--- while independent Landau analysis of the polarization gave $\beta=0.18\pm0.01$ (tricritical proximity) \cite{2026Yurtseven}. The framework's two-exponent prediction: Na-site channels matching the polar irrep grow as $(T_c-T)^{0.18}$---the same exponent as $P_s$, being realizations of the same irrep---while $A_1$ channels shift as
$(T_c-T)^{0.36}$. Re-extraction of both exponents from the 1978 tensor data, under the fitting protocol of SM Sec.~S9, is proposed here.

\subsection{Symmetry predictions for first-principles calculations}\label{sec:dft}

\subsection{Proposed case study: the $^{75}$As EFG as a nematic
order-parameter realization in BaFe$_2$As$_2$}\label{sec:case}

\emph{Choice.} This is the $p=1$ case; the alternative celebrated nematic,
FeSe, is disqualified for the centerpiece because $^{77}$Se is spin-1/2:
its NMR splitting is a Knight-shift observable, not an EFG one, while
$^{75}$As ($I=3/2$) is the quadrupolar workhorse of the 122 family. BaFe$_2$As$_2$ provides an ideal benchmark because its nematic transition is among the best-characterized examples of spontaneous symmetry breaking investigated by quadrupolar spectroscopy. The availability of high-quality crystallographic data, well-established NMR measurements, and a clearly identified transition irreducible representation allows every step of the present formalism, from site-symmetry decomposition to Landau coupling, to be demonstrated explicitly within a single material.

\emph{Structure and site symmetry.} Parent $I4/mmm$; As at $4e$
$(0,0,z_{\mathrm{As}})$, $z_{\mathrm{As}}\approx0.354$
\cite{2012Dhital}; $T_s\approx134$~K to $Fmmm$, orthorhombic axes at
$45^\circ$ to the tetragonal ones. Operations fixing the site: $E,2C_4,C_2,2\sigma_v,2\sigma_d$; inversion does not; hence $G_{\mathbf q}^{(0)}=C_{4v}$, non-centrosymmetric although
the space group is centrosymmetric. Orbit: two As per primitive cell at
$(0,0,\pm z)$, exchanged by inversion; index check
$|D_{4h}|/|C_{4v}|=2$ = orbit size.

\emph{Decomposition.} Characters $(5,-1,1,1,1)$ on
$(E,2C_4,C_2,2\sigma_v,2\sigma_d)$ give
$n_{A_1}=n_{B_1}=n_{B_2}=n_E=1$, $n_{A_2}=0$: the working decomposition
Eq.~\eqref{eq:d4h}, with the parent phase allowing exactly the axial
$V_{zz}$, $\eta=0$, as observed \cite{2008Kitagawa}.

\emph{Transition irrep; convention warning.} The orthorhombic distortion
makes the tetragonal diagonals inequivalent: the shear is $\epsilon_{xy}$,
transforming as $B_{2g}$ of $D_{4h}$ in the crystallographic (2-Fe) cell
used throughout; the 1-Fe-cell literature interchanges
$B_{1g}\leftrightarrow B_{2g}$. Daughter site group $C_{2v}$ (diagonal
mirrors retained); recount $n_{A_1}(C_{2v})=2$; subduction gives
$B_2\!\downarrow=A_1$ while $B_1\!\downarrow=A_2$,
$E\!\downarrow=B_1\oplus B_2$: \emph{exactly $V_{xy}$ switches on}.

\emph{Orbit induction and a null prediction.} With
$v^{(1,2)}=V^{(1,2)}_{xy}$ and inversion swapping sites while acting
trivially on parity-even rank-2 components,
\begin{equation}
v_\pm=\tfrac{1}{\sqrt2}\!\left(v^{(1)}\pm v^{(2)}\right):\quad
v_+\sim B_{2g},\qquad v_-\sim B_{2u}.
\end{equation}
The in-phase combination is the order-parameter realization; the
out-of-phase partner must remain zero through $T_s$. \emph{Null
prediction:} the two As sublattices acquire equal $V_{xy}$---same sign,
same magnitude; any resolved sublattice EFG inequivalence below $T_s$
(beyond twinning) would falsify the $B_{2g}$ assignment of the transition
itself. To our knowledge this prediction has not been stated.

\emph{Free energy.} With nematic amplitude $\varphi$ ($B_{2g}$;
agnostic between orbital and spin-nematic microscopics), critical channel
$v\equiv v_+$, shear $\epsilon_6$, spectator $s\equiv\Delta V_{zz}$:
\begin{align}
F&=\frac{a_0(T{-}T_0)}{2}\varphi^2+\frac{b}{4}\varphi^4
 -\lambda_v\varphi v+\frac{\kappa_v}{2}v^2
 -\lambda_\epsilon\varphi\epsilon_6\nonumber\\
&\quad+\frac{C_{66}^0}{2}\epsilon_6^2-\mu\,\epsilon_6 v
 -g\,s\varphi^2+\frac{\kappa_s}{2}s^2,
\end{align}
whose minimization over $(v,\epsilon_6,s)$ gives
\begin{equation}
v=\frac{\lambda_vC_{66}^0+\mu\lambda_\epsilon}
        {\kappa_vC_{66}^0-\mu^2}\,\varphi\equiv\Lambda\varphi,\qquad
s=\frac{g}{\kappa_s}\varphi^2,
\label{eq:Lambda}
\end{equation}
with $\Lambda$ decomposing into direct and strain-mediated paths
(separable by clamped-vs-relaxed DFT), and, above $T_s$,
$\partial v/\partial\epsilon_6\propto(T-T^*)^{-1}$: strain-NQR for a
specific nucleus.

\emph{Dictionary.} With only $V_{zz}$ and $V_{xy}$ nonzero, the in-plane
block diagonalizes by a $45^\circ$ rotation to eigenvalues
$-V_{zz}/2\pm V_{xy}$; the PAS keeps $Z\parallel c$ in the small-distortion
regime $\eta\ll1$ relevant at realistic orthorhombicity and
\begin{equation}
\eta=\frac{|2V_{xy}|}{|V_{zz}|}\propto|\varphi|,\qquad
\Delta\nu_Q=\nu_Q^0\!\left[\frac{\eta^2}{6}
 +\frac{\Delta V_{zz}}{V_{zz}^0}\right]+\mathcal O(4).
\end{equation}
Both powder terms are $\propto\varphi^2$ (the masquerade, concretely); the
clean observables are oriented-crystal satellite splittings along the two
orthorhombic axes, reading $2V_{xy}$ linearly. \emph{Prediction table:}
$\eta\propto(T_s{-}T)^\beta$; satellite anisotropy $\propto(T_s{-}T)^\beta$;
powder $\Delta\nu_Q\propto(T_s{-}T)^{2\beta}$; $v_-\equiv0$; Curie--Weiss
$dv/d\epsilon_6$ above $T_s$. Five signatures, one null.

\emph{Experiment.} The As-site EFG changes dramatically at the nematic
transition, with the two nematic domains interchanging the in-plane axes
\cite{2015Wiecki}---the domain-resolution bonus realized. In the
parent compound the magnetic transition follows immediately and the
combined transition is first-order-like, cutting off the asymptotic
critical regime; the division of labor is parent compound for
symmetry-level signatures and a separated-$T_s$ composition for the
exponent fit.

\emph{First principles (proposed).} We find no published channel-resolved
$V_{xy}(\delta)$ scan; the same frozen-mode parity test demonstrated on quartz in Sec.~\ref{sec:dft-quartz} applies to the $^{75}$As $B_{2g}$ channel; performing it here is left as the concrete next calculation.

\begin{figure*}[t]
\centering
\includegraphics[width=0.8\textwidth]{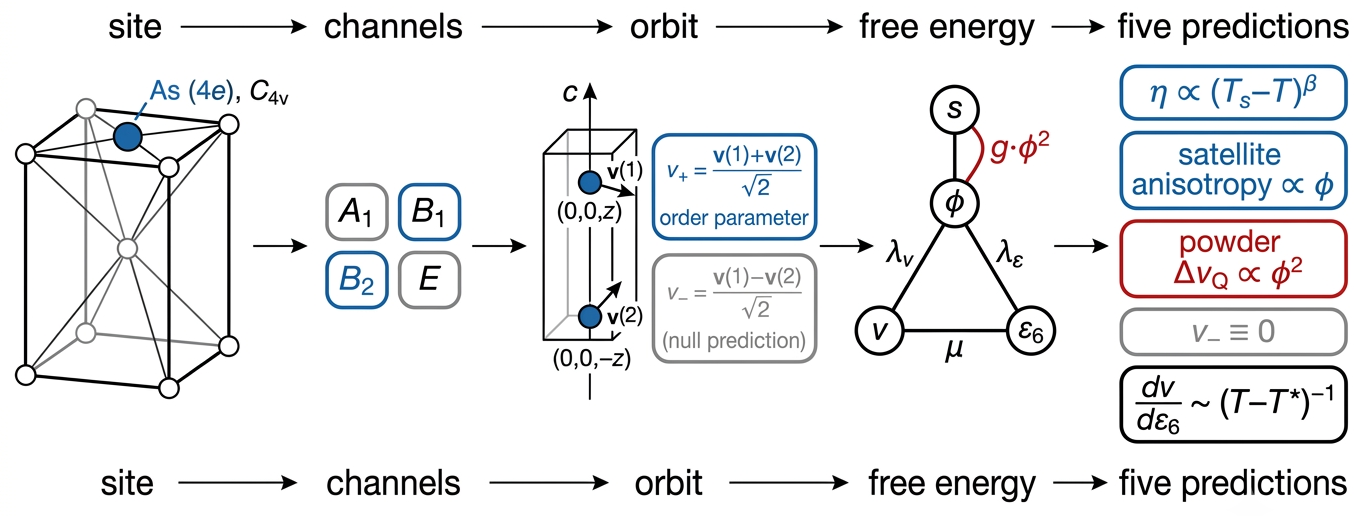}
\caption{The complete argument on one material (Sec.~\ref{sec:case}).
Left to right: the $^{75}$As site of BaFe$_2$As$_2$ and its $C_{4v}$ site
group; the channel decomposition with the $B_2$ ($V_{xy}$) channel
matching the transition irrep; induction over the two-site orbit, yielding
the order-parameter realization $v_+$ ($B_{2g}$) and the null-predicted
partner $v_-$ ($B_{2u}$); the invariant couplings among nematic amplitude
$\varphi$, EFG channel $v$, shear $\epsilon_6$, and the secondary channel
$s$; and the resulting five falsifiable predictions, color-coded by Landau
class (blue $p=1$, red $p=2$, gray forbidden).}
\label{fig:casestudy}
\end{figure*}

\emph{Assembled.} The As site symmetry forbids $V_{xy}$ in the parent
phase; the transition irrep $B_{2g}$ matches the site's $V_{xy}$ channel
through the orbit's in-phase combination; the Landau free energy forces
$V_{xy}\propto\varphi$ linearly and $\Delta V_{zz}\propto\varphi^2$, with
a strain-mediated decomposition of the coupling; the $^{75}$As dictionary
converts these into five falsifiable signatures including one null; the
experimental record confirms the symmetry-level signatures, with the
exponent test assigned to separated-$T_s$ compositions; and a small
first-principles campaign supplies the microscopic coefficients and parity
verification.

\section{Discussion}

\subsection{Implications}\label{sec:implications}

\emph{(i) The background-free channel.} Non-$A_1$ EFG channels have no
regular part to subtract: their onset is a symmetry identity. Among
standard order-parameter realizations this property is rare; for weak
transitions, where the anomaly-on-background problem dominates error
budgets, the symmetry-forbidden EFG channel is arguably the optimal
order-parameter observable available to local spectroscopy.
\emph{(ii) Site-multiplexed thermodynamics.} One transition makes
correlated predictions at every site whose induced representation contains
$\Gamma_\varphi$: common $T_c$, common $\beta$, site-dependent
$\Lambda^{(a)}$---an internally cross-checked measurement of one
thermodynamic function, with the ratios computable in advance.
\emph{(iii) The calibration loop.} DFT measures $v(\delta)$ Sec.~\ref{sec:dft-quartz}) and, at a transition, $v(\varphi)$; experiment measures $v(T)$; their composition determines $\varphi(T)$ from hyperfine data alone---subject to the quantitative-trust caveat of
Sec.~\ref{sec:limitations}.

\subsection{New readings of hyperfine data}\label{sec:readings}

The linear/quadratic relation becomes assignable, not assumed: historical exponents can be reclassified among secondary channels, masqueraded primary channels, and genuine tricriticality, disentangled by site analysis plus one oriented-crystal measurement. Null results become exclusion statements: a complete set of nulls generically excludes all irreps in the induced representation---the exclusionary logic that representation analysis gives neutron diffraction, now for hyperfine spectroscopy. Domain populations and wall dynamics, routinely discarded as twinning complications, are data: $p=1$ channels carry the sign of $\varphi$. For TDPAC, the framework transfers intact if the probe occupies the intended Wyckoff position; an aliovalent probe with bound defects locally lowers the site symmetry, and observation of a symmetry-forbidden
channel \emph{above} $T_c$ is then a fingerprint of probe-associated symmetry breaking---a nuisance converted into a diagnostic.

Existing analyses of quadrupolar observables typically proceed by postulating a phenomenological dependence of the EFG upon an independently defined order parameter. The present framework differs fundamentally in that the relevant EFG channel is identified directly from symmetry before any microscopic model is introduced. This distinction converts empirical correlations into symmetry-derived predictions and provides an explicit criterion for falsifiability.

\subsection{Limitations}\label{sec:limitations}

(i)~Zone-center scope (A4): zone-boundary and incommensurate transitions require extension. (ii)~Realization, not driver (A5, A6): where $\lambda$ is accidentally small the realization is experimentally faint; multi-site redundancy is the guard---and the reason null-result exclusions in Sec.~\ref{sec:readings} are generic rather than absolute.
(iii)~Statics only: symmetry zeros constrain means, not variances; fluctuation observables await the dynamic extension.
(iv)~Mean-field free energies; universality classes are borrowed, not derived. (v)~Quantitative trust in DFT amplitudes inherits the known systematics of first-principles EFGs; parity/zero structure is functional-independent, the numbers $\Lambda$ are not. (vi)~The EFG is time-even; magnetic order enters only through time-even composites.
(v) Rb$_2$ZnBr$_4$ (normal$\to$incommensurate) lies outside the zone-center scope (A4) but illustrates the dichotomy in modulated form: the local frequency carries linear and quadratic terms in the modulation amplitude, site symmetry dictating which survives (SM, Sec.~S9). $^{87}$Rb satellite NMR gives $\beta=0.35\pm0.01$ (central line) and $2\bar\beta=0.83\pm0.03$ (satellite splitting) \cite{1982Schneider,1982Blinc}. A full incommensurate treatment is deferred.

\subsection{Future directions}\label{sec:future}

\emph{(i) Landau--Ginzburg extension.} For the $D_{4h}$ $B_{2g}$ channel
the gradient theory has two stiffnesses,
$c_\parallel[(\partial_xv)^2+(\partial_yv)^2]+c_z(\partial_zv)^2$. A
domain wall $v(x)=v_0\tanh(x/\xi)$ then predicts a frequency histogram
\begin{equation}
\frac{dN}{dv}\propto\frac{1}{1-(v/v_0)^2},
\end{equation}
minimal at the wall center and edge-singular at the domain frequencies: van Hove-type singularities at $\pm\nu(v_0)$ with a shallow inter-domain continuum---a fittable prediction for wall-rich samples, and the local-spectroscopy analogue of diffuse scattering. \emph{(ii) Dynamics:} fluctuation spectra decompose by the same channels; the dichotomy reappears as the selection rule for critical $T_1$ divergence. \emph{(iii) The time-odd counterpart:} the hyperfine magnetic field is the axial, time-odd partner object; subduction onto magnetic site groups yields the parallel theorem, with the EFG entering magnetic transitions only at second order---the symmetry derivation of the weak
quadrupolar anomalies at purely magnetic transitions. \emph{(iv) Coupling lattice:} superconductivity couples through $|\Delta|^2$ (time-even, gauge-invariant), predicting small, symmetry-classifiable $\nu_Q$ anomalies at $T_c$; the conjugate-field program makes elasto-NQR the hyperfine sibling of elastoresistance. \emph{(v) Tabulation:} the per-(space group, Wyckoff) induced-representation table---the EFG analogue
of the spontaneous-strain tables---is mechanical to generate and would make the framework a lookup tool.

\emph{Relation to established representation-analysis methods.} The group-theoretical machinery used here---subduction of a physical-quantity space onto site groups, induction over Wyckoff orbits, and classification by parent-group irreps---is standard, and underlies symmetry-mode analysis of displacive transitions \cite{1972Bradley}, its computational implementations (e.g.\ the {\sc isotropy} and Bilbao {\sc amplimodes} tools \cite{ISOTROPY,AMPLIMODES}), and the spontaneous-strain theory of ferroelastics \cite{1990Salje,1998Carpenter}. Our contribution is not a new formalism but the identification of the nuclear-site EFG as a carrier space to which this machinery applies, together with the consequences specific to that choice. Three points distinguish the EFG case from displacement- and strain-mode analysis. First, the carrier is a rank-2, purely $\ell=2$ tensor \emph{localized at a crystallographic site}, so the relevant object is the site group rather than the full unit-cell displacement pattern, and the induction step---not the single-site subduction---is what supplies the parent-group order-parameter realization. Second, because the EFG is directly and site-selectively measured by hyperfine spectroscopy, the abstract irrep labels acquire an immediate experimental reading: the linear/quadratic dichotomy of Eq.~\eqref{eq:dichotomy} becomes an assignable critical exponent rather than a bookkeeping device. Third, the symmetry-forbidden EFG channels are \emph{background-free} by the vanishing identity, a property displacement modes (which sit on a nonzero reference structure) do not share, and which underlies the null-result exclusion logic. In this sense the present framework is to hyperfine spectroscopy what symmetry-mode analysis is to diffraction: the same representation theory, applied to a different, complementary observable.

\section{Conclusions}

We have shown that the electric field gradient tensor at a nuclear
site---long employed as an empirical proxy for order parameters defined
elsewhere---is itself, under precisely stated conditions, a
symmetry-adapted realization of the order parameter of a structural or
electronic phase transition. The argument rests on three exact structural
facts: the EFG is a pure $\ell=2$ object---traceless and symmetric not by
idealization but by the exact decoupling of the trace from every
quadrupolar observable; it decomposes under the site-symmetry group and
assembles, by induction over the Wyckoff orbit, into irreducible
representations of the parent group; and whenever the transition's irrep
appears in that decomposition, the corresponding orbit-adapted EFG
combination vanishes identically above $T_c$, couples linearly to the
order parameter, and inherits its critical behavior---exponent, sign, and
domain structure included. The complementary channels grow quadratically,
which derives, rather than assumes, the classic
$\Delta\nu_Q\propto\varphi^2$ phenomenology. The resulting
classification---primary, secondary, or forbidden, computable from the
space group, Wyckoff position, and transition channel alone---is the
paper's central deliverable.

The framework finds strong support from existing experimental observations and first-principles calculations, which consistently reproduce the symmetry relations derived here. Although further experimental tests of the newly proposed null predictions remain desirable, the available evidence strongly suggests that the EFG constitutes a robust symmetry-adapted realization of structural order parameters across a broad class of continuous phase transitions.

The claim's boundary is part of the claim: the EFG realizes order
parameters; it does not drive transitions. The construction covers
zone-center transitions of time-even order parameters, constrains means
rather than fluctuations, and borrows its universality classes. Each
boundary is a direction: the Ginzburg extension puts domain walls into the
EFG lineshape, the dynamic extension brings relaxation into scope, and the
time-odd counterpart---the hyperfine magnetic field as the realization of
magnetic order parameters---completes a program in which the two hyperfine
observables measured by a single TDPAC or M\"ossbauer experiment become
the time-even and time-odd order-parameter realizations of one transition.

To the Landau-theory tradition: the inventory of order-parameter carriers
has a further member, measured site-selectively at parts-per-million
precision. To hyperfine spectroscopy: the relation between quadrupolar
observables and order parameters, empirical since the founding of the
field, is now a theorem with an assignment rule, a falsifiability clause,
and a null-result logic. And to first-principles practice: the
coefficients of the Landau free energy in EFG variables are exactly the
derivatives that frozen-distortion calculations compute, making hyperfine
DFT and Landau phenomenology mutually calibrating---a loop through which
the order parameter of a transition can be determined from hyperfine data
alone.

\begin{acknowledgments}
    This research was funded by the Fundação de Amparo à Pesquisa do Estado de São Paulo (FAPESP) through grant numbers 2014/14001-1 and 2017/50332-0. A.W.C.  acknowledges the Conselho Nacional de Desenvolvimento Científico e Tecnológico (CNPq) for financial support through grant numbers 307322/2021-1, 408139/2022-6, 404949/2024-0, and 445223/2024-3. Additionally, this work was supported by the Coordenação de Aperfeiçoamento de Pessoal de Nível Superior (CAPES) through a scholarship.
    
\end{acknowledgments}

\section*{Declaration of Interest}
The authors declare no competing interests.

\section*{Declaration of generative AI and AI-assisted technologies in the writing process}

During the preparation of this work, the author(s) used ChatGPT and Perplexity in order to search for references and also check the grammar and improve the writting. After using this tool/service, the author(s) reviewed and edited the content as needed and take(s) full responsibility for the content of the published article.

\bibliographystyle{unsrt}
\bibliography{sample}

\clearpage
\onecolumngrid
\section*{Supplementary Materials}

This Supplemental Material collects the derivations, tables, and protocols
referenced in the main text. Section~\ref{sm:generators} gives the explicit
action of the site-symmetry generators on the five electric field gradient
(EFG) components. Section~\ref{sm:decomp} tabulates the subduction of the
$\ell=2$ representation over the crystallographic point groups, from which the
vanishing theorem follows. Section~\ref{sm:projector} verifies the projection
operators used to build the symmetry-adapted basis. Section~\ref{sm:pas}
collects the tensor-transformation chain rule and the principal-axis-system
(PAS) algebra, including the bound $0\le\eta\le1$. Section~\ref{sm:invariants}
develops the invariant algebra of the traceless symmetric tensor, the change of
order-parameter realization, and the multidimensional channels.
Section~\ref{sm:uniaxial} proves the stability of the uniaxial section.
Section~\ref{sm:firstorder} solves the cubic-invariant and tricritical free
energies. Section~\ref{sm:observables} records the spin-dependent observable
conventions. Section~\ref{sm:protocols} states the exponent-fitting and
frozen-mode protocols and derives the linear/quadratic structure in modulated
phases. Equation, section, and assumption labels (A1--A12) refer to the main
text unless prefixed by ``S''.

\section{Generator matrices and their action on the EFG components}
\label{sm:generators}

The five independent EFG components may be organized as
$\{V_{zz},\,V_{xx}-V_{yy},\,V_{xy},\,V_{xz},\,V_{yz}\}$. For each point-group
generator $g$ with proper rotation matrix $R(g)$, the induced action on the
tensor follows from $V'_{ij}=R_{ik}R_{jl}V_{kl}$, using $V_{ij}=V_{ji}$
throughout. Because $R_{ik}R_{jl}$ is even under $R\to-R$, every improper
operation acts identically to its rotational skeleton; it therefore suffices to
tabulate the proper generators and inversion. This reproduces the main-text
action of $C_{4z}$ and supplies the remaining generators used in the
subductions of Sec.~\ref{sm:decomp}.

\paragraph*{$C_{4z}$ (rotation angle $\pi/2$).}
With $R=\begin{pmatrix}0&-1&0\\ 1&0&0\\ 0&0&1\end{pmatrix}$, a
component-by-component evaluation gives
\begin{align}
V'_{xx}&=R_{1k}R_{1l}V_{kl}=R_{12}R_{12}\,V_{yy}=V_{yy}, \nonumber\\
V'_{yy}&=V_{xx},\qquad V'_{zz}=V_{zz}, \nonumber\\
V'_{xy}&=R_{12}R_{21}\,V_{yx}=-V_{xy}, \nonumber\\
V'_{xz}&=R_{12}R_{33}\,V_{yz}=-V_{yz},\qquad
V'_{yz}=R_{21}R_{33}\,V_{xz}=+V_{xz}.
\end{align}
Collected in the symmetry-adapted components,
\begin{equation}
V_{zz}\!\to\!V_{zz},\quad
(V_{xx}{-}V_{yy})\!\to\!-(V_{xx}{-}V_{yy}),\quad
V_{xy}\!\to\!-V_{xy},\quad
(V_{xz},V_{yz})\!\to\!(-V_{yz},V_{xz}),
\end{equation}
which is the main-text action.

\paragraph*{Remaining generators.} The same procedure yields:
\begin{itemize}
\item \emph{$C_{2z}$} $[R=\mathrm{diag}(-1,-1,1)]$:
$V_{zz},(V_{xx}{-}V_{yy}),V_{xy}$ invariant;
$(V_{xz},V_{yz})\to(-V_{xz},-V_{yz})$.
\item \emph{$C_{2x}$} $[R=\mathrm{diag}(1,-1,-1)]$:
$V_{zz}$ and $(V_{xx}{-}V_{yy})$ invariant; $V_{xy}\to-V_{xy}$;
$V_{xz}\to-V_{xz}$; $V_{yz}\to+V_{yz}$.
\item \emph{$\sigma_v(xz)$} $[y\to-y]$:
$V_{xy}\to-V_{xy}$; $V_{yz}\to-V_{yz}$; the remaining three invariant.
\item \emph{$\sigma_d$} $[x\leftrightarrow y]$:
$V_{xy}\to+V_{xy}$; $(V_{xx}{-}V_{yy})\to-(V_{xx}{-}V_{yy})$;
$(V_{xz},V_{yz})\to(V_{yz},V_{xz})$.
\item \emph{$C_{3z}$} $[\phi=2\pi/3]$: in the complex combinations
$u_{\pm2}=(V_{xx}{-}V_{yy})\pm 2iV_{xy}$ and $u_{\pm1}=V_{xz}\pm iV_{yz}$,
one has $u_{m}\to e^{-im\phi}u_{m}$ with $V_{zz}$ invariant; the integer $m$ is
the azimuthal weight, exhibiting the $\ell=2$ multiplet structure directly.
\item \emph{Inversion} $[R=-\mathbb{1}]$: all five components invariant, since
the EFG is parity-even.
\end{itemize}

These entries determine, for any site group generated by the operations above,
which of the five components survive projection onto the identity irrep
(Sec.~\ref{sm:decomp}) and how the remaining components are permuted among
themselves.

\section{Subduction of $D^{(\ell=2)}$ over the crystallographic point groups}
\label{sm:decomp}

The equilibrium EFG is the projection of the $\ell=2$ carrier space onto the
identity irrep $A_1$ of the site group $G_{\mathbf q}$; the number of
symmetry-allowed components is the multiplicity
\begin{equation}
n_{A_1}=\frac{1}{|G_{\mathbf q}|}\sum_{g\in G_{\mathbf q}}
\chi^{(\ell=2)}(g),\qquad
\chi^{(\ell=2)}(\phi)=1+2\cos\phi+2\cos 2\phi,
\label{eq:sm-counting}
\end{equation}
with $\phi$ the rotation angle of $g$ and improper classes evaluated on their
rotational skeleton (the EFG being parity-even). The full subduction
$D^{(\ell=2)}\!\downarrow G_{\mathbf q}=\bigoplus_\mu n_\mu\,\Gamma^{(\mu)}$
follows from $n_\mu=|G_{\mathbf q}|^{-1}\sum_g\chi^{(\mu)}(g)^*\chi^{(\ell=2)}(g)$.
Table~\ref{tab:sm-decomp} lists the results grouped by rotational skeleton,
each row satisfying the checksum $\sum_\mu n_\mu d_\mu=5$
($d_\mu$ = irrep dimension). Groups sharing a rotational skeleton
(e.g.\ $C_{4v},D_4,D_{2d},D_{4h}$) share the subduction, since the EFG is
insensitive to the proper/improper distinction.

\begin{table}[h]
\caption{Subduction of the $\ell=2$ (EFG) representation onto the
crystallographic point groups, and the number $n_{A_1}$ of symmetry-allowed
(equilibrium) components from Eq.~\eqref{eq:sm-counting}. Parenthetical
subscripts $(g)$ indicate the gerade label for centrosymmetric groups. Complex
one-dimensional irreps of the cyclic groups combine pairwise into physically
irreducible (real) doublets, indicated by the paired listing. Every row obeys
the checksum $\sum_\mu n_\mu d_\mu=5$. The equilibrium EFG vanishes
($n_{A_1}=0$) iff the site class is cubic.}
\label{tab:sm-decomp}
\begin{ruledtabular}
\begin{tabular}{lll}
Site group & $D^{(\ell=2)}$ subduction & $n_{A_1}$ \\ \hline
$C_1,\ C_i$ & $5A_{(g)}$ & 5 \\
$C_2,\ C_s,\ C_{2h}$ & $3A_{(g)}\oplus 2B_{(g)}$ & 3 \\
$C_{2v}$ & $2A_1\oplus A_2\oplus B_1\oplus B_2$ & 2 \\
$D_2,\ D_{2h}$ & $2A_{(g)}\oplus B_{1(g)}\oplus B_{2(g)}\oplus B_{3(g)}$ & 2 \\
$C_4,\ S_4,\ C_{4h}$ & $A_{(g)}\oplus 2B_{(g)}\oplus E_{(g)}$ & 1 \\
$C_{4v},\ D_4,\ D_{2d},\ D_{4h}$ &
  $A_{1(g)}\oplus B_{1(g)}\oplus B_{2(g)}\oplus E_{(g)}$ & 1 \\
$C_3,\ S_6$ & $A_{(g)}\oplus 2E_{(g)}$ & 1 \\
$C_{3v},\ D_3,\ D_{3d}$ & $A_1\oplus 2E_{(g)}$ & 1 \\
$C_6,\ C_{3h},\ C_{6h}$ & $A_{(g)}\oplus E_{1(g)}\oplus E_{2(g)}$ & 1 \\
$C_{6v},\ D_6,\ D_{3h},\ D_{6h}$ &
  $A_{1(g)}\oplus E_{1(g)}\oplus E_{2(g)}$ & 1 \\
$T,\ T_h$ & $E_{(g)}\oplus T_{(g)}$ & 0 \\
$T_d,\ O,\ O_h$ & $E_{(g)}\oplus T_{2(g)}$ & 0 \\
\end{tabular}
\end{ruledtabular}
\end{table}

\paragraph*{Cubic vanishing theorem (worked case $O_h$).}
Evaluating Eq.~\eqref{eq:sm-counting} on the classes
$(E,8C_3,6C_2,6C_4,3C_2)$ with $\chi^{(\ell=2)}=(5,-1,1,-1,1)$,
\begin{equation}
n_{A_{1g}}(O_h)=\frac{1}{48}\big[\,5-8+6-6+3\,\big]\times 2 = 0,
\qquad D^{(\ell=2)}\!\downarrow O_h=E_g\oplus T_{2g},
\end{equation}
where the factor of $2$ accounts for the improper classes of identical
rotational skeleton. The same computation on $T$ and $T_h$ gives
$E\oplus T$ with $n_{A_1}=0$. For every non-cubic group in
Table~\ref{tab:sm-decomp}, $n_{A_1}\ge 1$. Hence the equilibrium EFG vanishes
identically if and only if the site class is cubic, the sharp statement quoted
in the main text.

\paragraph*{Worked orbit case: the $4e$ position of $I4/mmm$.}
For the $^{75}$As case study the site group is $C_{4v}$, giving the working
decomposition
$D^{(\ell=2)}\!\downarrow C_{4v}=A_1\oplus B_1\oplus B_2\oplus E$ with the
assignment $A_1\leftrightarrow V_{zz}$, $B_1\leftrightarrow V_{xx}-V_{yy}$,
$B_2\leftrightarrow V_{xy}$, $E\leftrightarrow(V_{xz},V_{yz})$. In the parent
phase only $V_{zz}$ ($\eta=0$) is allowed, and the $B_{2g}$ nematic distortion
activates $V_{xy}$ (main text). The orbit contains two sites exchanged by the
lost inversion; the in-phase and out-of-phase combinations transform as
$B_{2g}$ and $B_{2u}$ respectively, which is the induction result used to
predict the sublattice null.

\section{Verification of the projection operators}
\label{sm:projector}

The symmetry-adapted basis is generated by the projectors
$\hat P^{(\mu)}_{mm}=(d_\mu/|G_0|)\sum_g [D^{(\mu)}_{mm}(g)]^*\hat O(g)$.
We verify the two properties used in the main text: that
$\hat P^{(\mu)}_{mm}$ projects onto row $m$ of $\Gamma^{(\mu)}$, and that the
projectors resolve the identity.

\paragraph*{Transformation property.}
Acting with a group element $\hat O(h)$ and relabelling $g'=hg$ (the
rearrangement theorem lets the sum run over $g'$),
\begin{align}
\hat O(h)\,\hat P^{(\mu)}_{mm}
&=\frac{d_\mu}{|G_0|}\sum_{g}[D^{(\mu)}_{mm}(g)]^*\,\hat O(hg)
 =\frac{d_\mu}{|G_0|}\sum_{g'}[D^{(\mu)}_{mm}(h^{-1}g')]^*\,\hat O(g').
\end{align}
Using the homomorphism $D^{(\mu)}(h^{-1}g')=D^{(\mu)}(h^{-1})D^{(\mu)}(g')$ and
unitarity $D^{(\mu)}_{mm}(h^{-1})=[D^{(\mu)}_{mm}(h)]^*$ together with
$[D^{(\mu)}_{mn}(g')]^*$ closure,
\begin{equation}
\hat O(h)\,\hat P^{(\mu)}_{mm}
=\sum_{n} D^{(\mu)}_{nm}(h)\,\hat P^{(\mu)}_{nm},
\end{equation}
so $\hat P^{(\mu)}_{mm}\phi$ transforms as row $m$ of $\Gamma^{(\mu)}$, as
required. 

\paragraph*{Completeness.}
Summing the diagonal projectors over irreps and rows and using the great
orthogonality theorem $\sum_\mu (d_\mu/|G_0|)\sum_g \chi^{(\mu)}(g)^*
\delta_{\cdots}=\mathbb{1}$ (equivalently, character completeness
$\sum_\mu d_\mu\chi^{(\mu)}(g)=|G_0|\,\delta_{g,e}$) gives
\begin{equation}
\sum_{\mu}\sum_{m}\hat P^{(\mu)}_{mm}=\mathbb{1}.
\end{equation}
The two properties together guarantee that projecting the five-component EFG
space yields a complete, non-overlapping symmetry-adapted basis, which is the
construction used in Sec.~III of the main text.

\section{Tensor chain rule, PAS algebra, and the bound $0\le\eta\le1$}
\label{sm:pas}

\paragraph*{Chain rule.}
Under a rotation $x'_i=R_{ij}x_j$ with $R$ orthogonal, the second derivatives
transform as a rank-2 tensor:
\begin{equation}
\frac{\partial}{\partial x'_i}
=\frac{\partial x_k}{\partial x'_i}\frac{\partial}{\partial x_k}
=R_{ik}\frac{\partial}{\partial x_k}
\;\Longrightarrow\;
V'_{ij}=\frac{\partial^2 V}{\partial x'_i\partial x'_j}
=R_{ik}R_{jl}\,V_{kl},
\end{equation}
using $\partial x_k/\partial x'_i=R_{ik}$ from orthogonality. This is the
transformation law quoted in the main text; the trace
$\mathrm{Tr}\,V'=R_{ik}R_{il}V_{kl}=\delta_{kl}V_{kl}=\mathrm{Tr}\,V$ is
invariant, consistent with the traceless projection being well defined in every
frame.

\paragraph*{PAS parametrization.}
In the principal-axis system the EFG is diagonal with
$V_{XX}+V_{YY}+V_{ZZ}=0$. With the asymmetry parameter
$\eta\equiv(V_{XX}-V_{YY})/V_{ZZ}$ and tracelessness,
\begin{equation}
V_{XX}=-\tfrac{V_{ZZ}}{2}(1-\eta),\qquad
V_{YY}=-\tfrac{V_{ZZ}}{2}(1+\eta),
\end{equation}
which reproduces the main-text expressions.

\paragraph*{Bound $0\le\eta\le1$.}
The ordering convention $|V_{ZZ}|\ge|V_{YY}|\ge|V_{XX}|$ fixes the range of
$\eta$. From the parametrization,
\begin{align}
|V_{YY}|\ge|V_{XX}| &\Longrightarrow |1+\eta|\ge|1-\eta|
\Longrightarrow \eta\ge 0, \\
|V_{ZZ}|\ge|V_{YY}| &\Longrightarrow 1\ge\tfrac12(1+\eta)
\Longrightarrow \eta\le 1,
\end{align}
so $0\le\eta\le1$. At the endpoint $\eta=1$ the labels $X$ and $Y$ become
interchangeable and the ordering convention reassigns axes; consequently every
small-distortion statement in the main text (e.g.\ the pnictide dictionary)
is made in the regime $\eta\ll1$, where the assignment $Z\parallel c$ is
stable.

\section{Invariant algebra, change of realization, and multidimensional
channels}
\label{sm:invariants}

\paragraph*{Integrity basis of a traceless symmetric tensor.}
For a symmetric $3\times3$ tensor $Q$ with $\mathrm{Tr}\,Q=e_1=0$, the power
sums $p_n=\mathrm{Tr}(Q^n)$ and elementary symmetric functions $e_n$ are
related by Newton's identities:
\begin{equation}
p_1=e_1=0,\qquad
p_2=e_1 p_1-2e_2=-2e_2,\qquad
p_3=e_1 p_2-e_2 p_1+3e_3=3e_3,
\end{equation}
so $e_2=-\tfrac12 p_2$ and $e_3=\tfrac13 p_3$. Since
$e_3=\det Q$, one has $\det Q=\tfrac13\mathrm{Tr}(Q^3)$. For the quartic power
sum,
\begin{equation}
p_4=e_1 p_3-e_2 p_2+e_3 p_1
=-e_2 p_2=\tfrac12 p_2^2
\;\Longrightarrow\;
\mathrm{Tr}(Q^4)=\tfrac12\big[\mathrm{Tr}(Q^2)\big]^2 .
\end{equation}
A numerical check with eigenvalues $(2,-1,-1)$ gives $p_2=6$, $p_3=6$,
$p_4=18=\tfrac12(6)^2$, and $\det Q=2=\tfrac13 p_3$. Hence the integrity basis
of the isotropic invariants is $\{\mathrm{Tr}(Q^2),\mathrm{Tr}(Q^3)\}$, and the
most general quartic Landau expansion is the three-term form quoted in the main
text with no independent quartic invariant.

\paragraph*{Spectroscopic form of the invariants.}
Substituting the PAS parametrization,
\begin{equation}
\mathrm{Tr}(Q^2)=\tfrac{V_{ZZ}^2}{2}\,(3+\eta^2),\qquad
\mathrm{Tr}(Q^3)=\tfrac{3}{4}\,V_{ZZ}^3\,(1-\eta^2),
\end{equation}
so the quadratic invariant is the rotationally invariant EFG magnitude, while
the cubic invariant is odd in $V_{ZZ}$ (selecting prolate versus oblate) and
vanishes at maximal biaxiality $\eta=1$; a nonzero cubic coefficient therefore
drives uniaxial order at onset (Sec.~\ref{sm:uniaxial}).

\paragraph*{Change of order-parameter realization.}
The main text writes the free energy alternately in the microscopic order
parameter $\varphi$ and in the critical EFG channel $v$. Starting from
\begin{equation}
F=\frac{a_0(T-T_c)}{2}\varphi^2+\frac{b}{4}\varphi^4
-\lambda\varphi v+\frac{\kappa}{2}v^2 ,
\label{eq:sm-twofield}
\end{equation}
minimizing over $v$ at fixed $\varphi$ gives $v^*=(\lambda/\kappa)\varphi$;
substituting back leaves a quartic in $\varphi$ with a downshifted quadratic
coefficient, i.e.\ the pseudo-proper renormalization
$T_c^{\mathrm{eff}}=T_c+\lambda^2/(a_0\kappa)$. Conversely, to express $F$ in
$v$, minimize over $\varphi$ at fixed $v$:
\begin{equation}
a_0(T-T_c)\varphi+b\varphi^3=\lambda v .
\end{equation}
For small $v$, inverting perturbatively,
\begin{equation}
\varphi=\frac{\lambda v}{a_0(T-T_c)}
-\frac{b\lambda^3 v^3}{[a_0(T-T_c)]^4}+\mathcal{O}(v^5),
\end{equation}
and substituting back reproduces
\begin{equation}
F(v)=\frac{\tilde a\,(T-T_c^{\mathrm{eff}})}{2}\,v^2
+\frac{\tilde b}{4}\,v^4,\qquad
\tilde a=a_0\Big(\frac{\kappa}{\lambda}\Big)^{2},\quad
\tilde b=b\Big(\frac{\kappa}{\lambda}\Big)^{4},
\end{equation}
at the orders retained; the higher-order term renormalizes $\tilde b$ only.
Because $\lambda\neq0$, the map $\varphi\leftrightarrow v$ is locally
invertible near $T_c$, so the substitution and the minimization agree and the
two descriptions are equivalent realizations of the same transition, as stated
in the main text.

\paragraph*{Multidimensional channels.}
For a two-dimensional irrep $(v_1,v_2)$ the quadratic invariant is
$v_1^2+v_2^2$ and the leading quartic invariant is $(v_1^2+v_2^2)^2$, augmented
where the group allows by an anisotropy invariant. For a trigonal $E$ channel a
cubic invariant exists,
\begin{equation}
\mathrm{Re}\big[(v_1+iv_2)^3\big]=v_1^3-3v_1 v_2^2,
\end{equation}
invariant under $C_3$ (the phase $e^{2\pi i}=1$) and under the vertical mirror
(complex conjugation). Its presence renders the transition generically
first-order with a three-state-Potts minima structure; this is the crystalline
descendant of the de~Gennes cubic term, obtained here from the site group
rather than assumed from $\mathrm{O}(3)$. This discharges the one-dimensional
restriction (Assumption A7) invoked in the theorem: the multidimensional case
carries an additional cubic invariant but does not alter the linear
$v\propto\varphi$ coupling of the matching channel.

\section{Stability of the uniaxial section}
\label{sm:uniaxial}

We justify Assumption A12: when the cubic invariant is present ($C\neq0$),
the free-energy extremum lies on the uniaxial section and biaxial fluctuations
are stable (restoring). Parametrize the traceless tensor by its magnitude $S$
and a biaxiality coordinate $\eta_b$, so that the two invariants read
\begin{equation}
\mathrm{Tr}(Q^2)=S^2\,(1+\tfrac13\eta_b^2)+\mathcal{O}(\eta_b^4),\qquad
\mathrm{Tr}(Q^3)=S^3\,(1-\eta_b^2)+\mathcal{O}(\eta_b^4),
\end{equation}
with $\eta_b=0$ the uniaxial section. Inserting these into
$F=\tfrac{A}{2}\mathrm{Tr}(Q^2)-\tfrac{C}{3}\mathrm{Tr}(Q^3)
+\tfrac{B}{4}[\mathrm{Tr}(Q^2)]^2$ and expanding to quadratic order in
$\eta_b$,
\begin{equation}
F(S,\eta_b)=F(S,0)
+\Big[\tfrac{A}{6}S^2+\tfrac{C}{3}S^3+\tfrac{B}{6}S^4\Big]\eta_b^2
+\mathcal{O}(\eta_b^4).
\end{equation}
At the uniaxial minimum $S=S_0>0$ selected by $C>0$ (prolate branch), the
stationarity condition $\partial_S F(S,0)=0$ reads
$A S_0-C S_0^2+B S_0^3=0$, i.e.\ $A+B S_0^2=C S_0$. The coefficient of
$\eta_b^2$ then becomes
\begin{equation}
\tfrac16 S_0^2\big(A+B S_0^2\big)+\tfrac13 C S_0^3
=\tfrac16 C S_0^3+\tfrac13 C S_0^3
=\tfrac{1}{2}C S_0^3 \;>\;0 ,
\end{equation}
so biaxial fluctuations are restoring whenever $C\neq0$: the uniaxial section
is locally stable, and only at $C=0$ does the biaxial direction become soft.
This is the statement used in the main text to reduce the tensor free energy
to the scalar $S$ theory of Sec.~\ref{sm:firstorder}.

\section{First-order and tricritical solutions}
\label{sm:firstorder}

\paragraph*{Cubic-invariant (first-order) solution.}
On the uniaxial section the free energy is
\begin{equation}
F(S)=\frac{\alpha}{2}S^2-\frac{\gamma}{3}S^3+\frac{\beta_4}{4}S^4 .
\end{equation}
A first-order transition occurs where a nonzero minimum becomes degenerate with
$S=0$, i.e.\ $F(S)=0$ and $F'(S)=0$ simultaneously at $S=S_{\rm jump}$.
Writing $2F/S^2=\alpha-\tfrac{2\gamma}{3}S+\tfrac{\beta_4}{2}S^2=0$
and $F'/S=\alpha-\gamma S+\beta_4 S^2=0$ and subtracting to eliminate $\alpha$,
\begin{equation}
\tfrac{\gamma}{3}S=\tfrac{\beta_4}{2}S^2
\;\Longrightarrow\;
S_{\rm jump}=\frac{2\gamma}{3\beta_4},
\end{equation}
whence, substituting back into $F'/S=0$,
\begin{equation}
\alpha_{\rm tr}=\gamma S_{\rm jump}-\beta_4 S_{\rm jump}^2
=\frac{2\gamma^2}{9\beta_4}.
\end{equation}
The metastability (spinodal) limits are the values of $\alpha$ at which the
nonzero extrema appear and disappear, i.e.\ the discriminant boundaries of
$\beta_4 S^2-\gamma S+\alpha=0$: the ordered minimum first appears at
$\gamma^2=4\alpha\beta_4$ and the disordered minimum loses stability at
$\alpha=0$. These reproduce the main-text results (discontinuous $V_{ZZ}$
onset, uniaxial at birth, hysteresis window of width
$\alpha_{\rm tr}=2\gamma^2/9\beta_4$).

\paragraph*{Tricritical solution.}
When symmetry forbids the cubic invariant and the quartic coefficient is tuned
through zero ($b\to0$), the leading stabilizing term is sextic:
\begin{equation}
F(\varphi)=\frac{a_0(T-T_c)}{2}\varphi^2+\frac{c_6}{6}\varphi^6 .
\end{equation}
Minimizing, $a_0(T-T_c)\varphi+c_6\varphi^5=0$ gives, below $T_c$,
\begin{equation}
\varphi=\Big[\frac{a_0(T_c-T)}{c_6}\Big]^{1/4}
\;\Longrightarrow\;
\beta_{\rm tric}=\tfrac14 .
\end{equation}
This is the tricritical mean-field exponent quoted in the discussion of
TTF--chloranil, where the measured $\beta\approx0.16<\tfrac14$ places the
system on the weakly first-order side of a tricritical point.

\section{Spin-dependent observable conventions}
\label{sm:observables}

\paragraph*{Pure NQR, $I=3/2$.}
In the PAS the quadrupole Hamiltonian is
\begin{equation}
H_Q=\frac{eQV_{ZZ}}{4I(2I-1)}
\Big[\,3I_z^2-I^2+\eta\,(I_x^2-I_y^2)\,\Big].
\end{equation}
For $I=3/2$ the operator couples $\{|{+}\tfrac32\rangle,|{-}\tfrac12\rangle\}$
and $\{|{-}\tfrac32\rangle,|{+}\tfrac12\rangle\}$; diagonalizing either
$2\times2$ block gives the doubly degenerate eigenvalues
$\pm\tfrac12\,h\nu_Q$ with the single resonance frequency
\begin{equation}
\nu_Q=\frac{eQV_{ZZ}}{2h}\sqrt{1+\frac{\eta^2}{3}} ,
\end{equation}
the expression quoted in the main text. The $\sqrt{1+\eta^2/3}$ factor is the
origin of the ``masquerade'': for a transverse critical channel appearing where
$\eta^{(0)}=0$, $\nu_Q\propto 1+\eta^2/6+\cdots$, so a $p=1$ EFG channel
enters a single powder line quadratically.

\paragraph*{Quadrupole-perturbed NMR satellites.}
To first order in the quadrupole interaction the satellite shift of the
$m\!\leftrightarrow\!m{-}1$ transition is
\begin{equation}
\Delta\nu_m=\frac{3eQ\,(2m-1)}{4I(2I-1)h}\,V_{z'z'},\qquad
V_{z'z'}=\hat z'\!\cdot V\cdot\hat z' ,
\end{equation}
linear in the laboratory-frame component $V_{z'z'}$ and hence a linear
functional of the Cartesian EFG components. Oriented-crystal satellite
spectroscopy therefore reads the EFG channels linearly, with no masquerade;
this is the basis of the strain-NQR and sublattice-null proposals in the main
text.

\paragraph*{TDPAC ($I=5/2$).}
The static quadrupole interaction splits the $I=5/2$ level into three
observable frequencies $\omega_1,\omega_2,\omega_3$ with
$\omega_1+\omega_2=\omega_3$ at $\eta=0$; for general $\eta$ they are the
differences of the eigenvalues of $H_Q$ and are conventionally parametrized by
$\omega_0=3\pi eQV_{ZZ}/[2I(2I-1)h]$ and $\eta$. The lowest observed frequency
$\omega_1(\omega_0,\eta)$ increases monotonically with $\eta$; standard tables
(and the exact secular equation) give the $\eta$-dependence used to convert
measured triples into $(V_{ZZ},\eta)$.

\paragraph*{M\"ossbauer.}
For a $\tfrac32\!\to\!\tfrac12$ transition (e.g.\ $^{57}$Fe) the quadrupole
splitting is
\begin{equation}
\Delta E_Q=\frac{eQV_{ZZ}}{2}\sqrt{1+\frac{\eta^2}{3}} ,
\end{equation}
the same $(V_{ZZ},\eta)$ combination as the NQR frequency, so the masquerade
caveat applies identically. In all three cases the same two shape scalars
$(V_{ZZ},\eta)$ are the measured quantities, and the linear/quadratic
distinction is resolved only by an orientation-resolved measurement.

\section{Protocols and the modulated linear/quadratic derivation}
\label{sm:protocols}

\paragraph*{Exponent-fitting protocol.}
(1) Classify the observable ($p=1$, $p=2$, or nonlinear-map) from the site
analysis \emph{before} fitting, and model it as $y=A\,(1-T/T_c)^{p\beta}$.
(2) Subtract the regular background only in $A_1$ channels: fit
$\nu_Q^{\rm reg}(T)$ above $T_c$ and extrapolate; symmetry-forbidden channels
have zero background by the vanishing identity and need no subtraction.
(3) Treat $T_c$ as a fit parameter and report the $\hat T_c$--$\hat\beta$
covariance, since these are strongly correlated. (4) Perform a range-of-fit
study $\hat\beta(t_{\min},t_{\max})$ to expose crossover or correction-to-scaling
drift. (5) Screen first-order transitions against a jump-plus-hysteresis model
before quoting an exponent.

\paragraph*{Frozen-mode DFT protocol.}
(1) Relax the parent under the symmetry constraint; verify that forbidden
channels sit at the numerical noise floor, and record that floor. (2) Define
the distortion by irrep projection of the daughter$-$parent difference (or the
soft-mode eigenvector), and project out same-irrep admixtures. (3) Scan
\emph{both} signs of the amplitude, including small values, so that parity can
be read directly. (4) Project each computed tensor onto the symmetry-adapted
channels \emph{before} fitting, because the PAS scalars $(V_{ZZ},\eta)$ mix
channels nonlinearly. (5) Fit odd channels to $c_1\delta+c_3\delta^3$ and even
channels to $v_0+c_2\delta^2+c_4\delta^4$, and check that the forbidden
channels vanish. (6) Repeat with strain relaxed versus clamped to separate the
direct and strain-mediated parts of the coupling. (7) Cross-check a
representative slope $c_1$ between an all-electron (FP-LAPW) and a
pseudopotential (PAW) calculation at the few-percent level. The
$\alpha$-quartz analysis of the main text follows steps (2)--(5) at the orbit
level; the single-signed scan there is the reason the finer single-site
parity/zero-tests are deferred to a two-signed pure-mode refinement.

\paragraph*{Linear/quadratic structure in modulated phases.}
In a one-dimensionally modulated (incommensurate) phase the local displacement
is $u(\mathbf r)=u_0\cos[\mathbf q\cdot\mathbf r+\phi(\mathbf r)]$, and the
local resonance frequency expands in the local order parameter as
\begin{equation}
\nu(\mathbf r)=\nu_0+V_1\,u(\mathbf r)+\tfrac12 V_2\,u(\mathbf r)^2+\cdots
\end{equation}
The site symmetry of the position in the \emph{average} structure fixes whether
the linear coefficient $V_1$ is allowed: if the site is invariant under an
operation that sends $u\to-u$, then $V_1=0$ and the leading response is
quadratic, $\nu-\nu_0\propto u^2$. This is the modulated form of the main-text
dichotomy: off-diagonal EFG components (odd under the relevant operation) carry
the $V_1$ (linear, $p=1$) term, whereas the diagonal components (even) carry
only $V_2$ (quadratic, $p=2$). The two cases produce distinguishable edge
structure in the frequency distribution: with the plane-wave density of states
$dN/d\nu\propto|d\nu/dx|^{-1}$, a purely linear local coupling gives a
two-edge lineshape symmetric about $\nu_0$, whereas a purely quadratic coupling
gives a one-edge lineshape bounded on one side by $\nu_0$. This is the $V_1/V_2$
structure of the incommensurate NMR literature and underlies the two-exponent
behavior ($\beta$ versus $2\beta$) quoted for Rb$_2$ZnBr$_4$ in the main text.

\section{Regression statistics for the model fits}
\label{sm:statistics}

\begin{table}[t]
\caption{Regression statistics for the odd-model fits $v=a\delta+b\delta^{3}$
of the Si $V_{xy}$ orbit combinations (real ELK dataset, 21 amplitudes).
Standard errors, $t$-statistics and two-sided $p$-values are from the
residual covariance. The cubic coefficient $b$ is statistically insignificant
in every case ($p>0.4$), so the response is linear, not merely odd. The final
column reports $\Delta$AIC between the odd basis $\{\delta,\delta^3\}$ and the
even control basis $\{\delta^2,\delta^4\}$; strongly negative values decisively
favour the odd (parity-$p{=}1$) model.}
\label{tab:regression}
\begin{ruledtabular}
\begin{tabular}{lccccc}
Combination & $a$ (a.u./\AA$\sqrt{\rm amu}$) & $t_a$ ($p_a$) & $b$ ($p_b$) & $R^2$ & $\Delta$AIC \\ \hline
$v_-$ (activated)  & $-4.811(3)\times10^{-2}$ & $-150$ ($10^{-30}$) & $+4.1\times10^{-3}$ ($0.72$) & $0.99987$ & $-187$ \\
$v_+$ (suppressed) & $-8.88(2)\times10^{-4}$  & $-54$ ($10^{-22}$)  & $+4.7\times10^{-4}$ ($0.42$) & $0.99895$ & $-144$ \\
$A$ (inert ref.)   & $+3.76(2)\times10^{-4}$  & $+16$ ($10^{-12}$)  & $-6.8\times10^{-4}$ ($0.42$) & $0.98692$ & $-91$ \\
\end{tabular}
\end{ruledtabular}
\end{table}

\begin{figure}[t]
\centering
\includegraphics[width=\columnwidth]{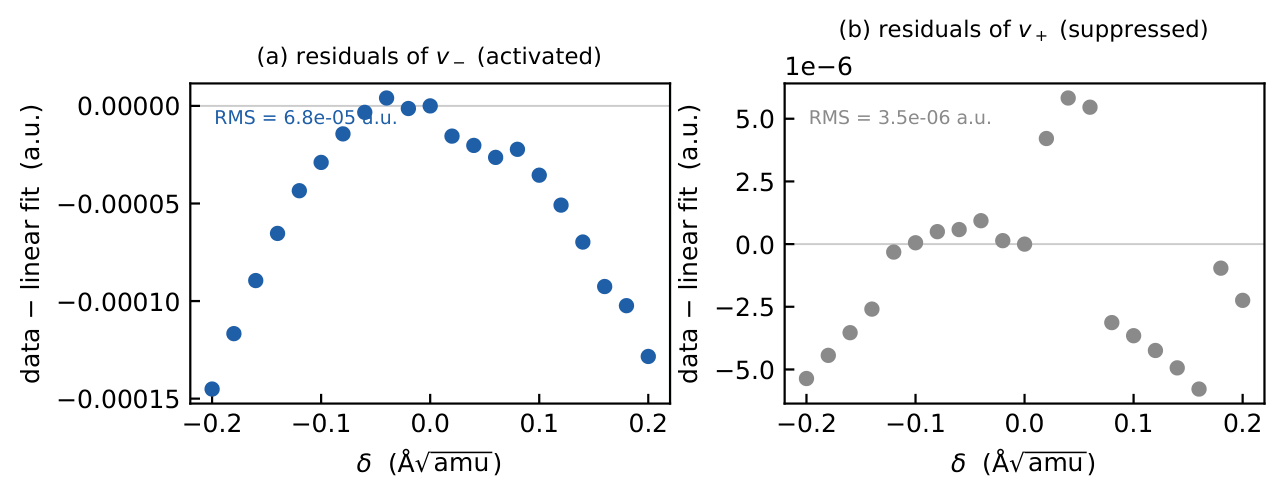}
\caption{Residuals of the linear model for the Si $V_{xy}$ orbit combinations
of $\alpha$-quartz, defined as the ELK data minus the fitted linear term
$a\delta$, versus the $A_1$ soft-mode amplitude $\delta$
(in $\mathrm{\AA}\sqrt{\mathrm{amu}}$). (a)~Activated combination
$v_-=(V_{xy}^{B}-V_{xy}^{C})/\sqrt{2}$: residuals remain below
$\sim\!1.5\times10^{-4}$~a.u.\ (RMS $6.8\times10^{-5}$~a.u.), two orders of
magnitude below the activated signal. (b)~Suppressed combination
$v_+=(V_{xy}^{B}+V_{xy}^{C})/\sqrt{2}$: residuals below
$\sim\!6\times10^{-6}$~a.u.\ (RMS $3.5\times10^{-6}$~a.u.; note the different
vertical scale). The residuals are not white noise but show a weak, smooth,
even-in-$\delta$ curvature; this is the expected imprint of the small
higher-$A_1$ mode admixture in the displacement pattern
(Sec.~\ref{sm:protocols}), and it does not affect the linear slope $a$ or the
$|a_+/a_-|=0.018$ suppression ratio.}
\label{fig:residuals}
\end{figure}

latex
\begin{figure}[t]
\centering
\includegraphics[width=\columnwidth]{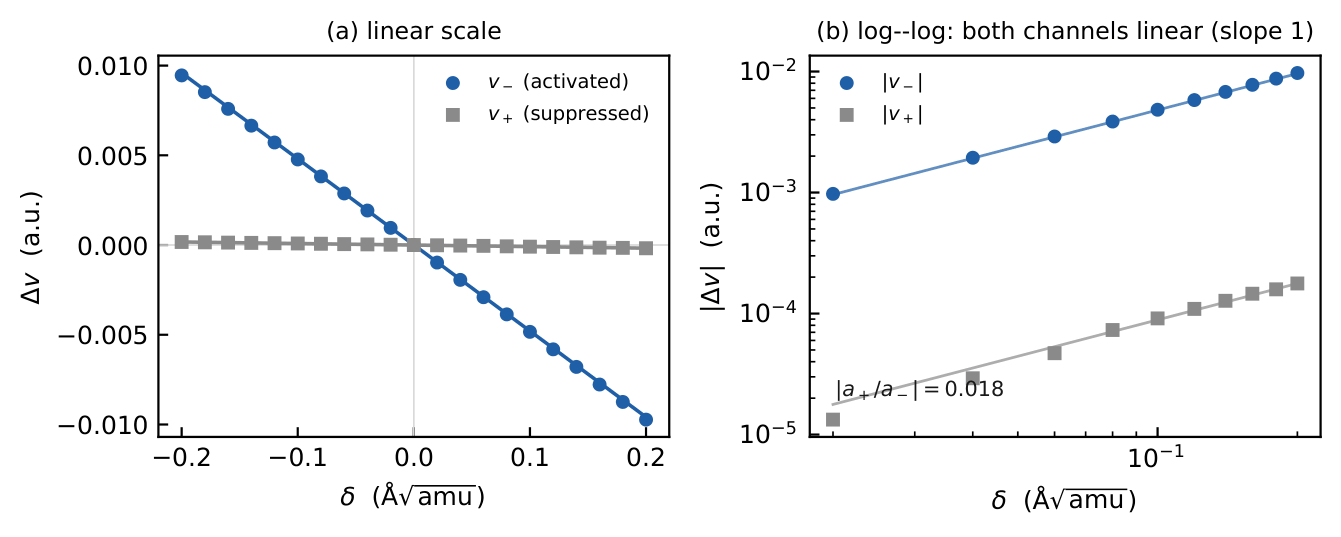}
\caption{Log-scale comparison of the activated and suppressed Si $V_{xy}$ orbit
combinations of $\alpha$-quartz, resolving the two channels across their
two-order-of-magnitude separation in amplitude. (a)~Linear scale: the activated
combination $v_-$ (circles, with the fitted $a\delta+b\delta^3$ curve) grows
steeply and linearly with the soft-mode amplitude $\delta$, while the suppressed
combination $v_+$ (squares) is nearly flat on the same axis. (b)~Log--log plot
of $|\Delta v|$ versus $\delta$ for $\delta>0$: both combinations track
slope-one reference lines, confirming that each is linear in $\delta$ and that
their difference is one of magnitude, not of power law; the offset between the
two lines is the suppression ratio $|a_+/a_-|=0.018$. EFG components are in
atomic units.}
\label{fig:logscale}
\end{figure}

\end{document}